\documentclass[preprints,article,accept,moreauthors]{Definitions/mdpi} 

\firstpage{1} 
\pubvolume{xx}
\issuenum{1}
\articlenumber{5}
\pubyear{2020}
\copyrightyear{2020}
\history{Received: date; Accepted: date; Published: date}

\usepackage[T1]{fontenc}
\usepackage{graphicx}
\usepackage{amsmath}
\newenvironment{sistema}%
{\left\lbrace\begin{array}{@{}l@{}}}%
{\end{array}\right.}

\Title{Caustics in gravitational lensing by mixed binary systems}

\Author{Valerio Bozza $^{1,2}*$, Silvia Pietroni $^{1,2}$ and Chiara Melchiorre $^{3}$}

\AuthorNames{Valerio Bozza, Silvia Pietroni and Chiara Melchiorre}

\address{%
$^{1}$ \quad Dipartimento di Fisica “E.R. Caianiello”, Università di Salerno, Via Giovanni Paolo II 132,
I-84084 Fisciano, Italy\\
$^{2}$ \quad  Istituto Nazionale di Fisica Nucleare, Sezione di Napoli, Via Cintia, 80126, Napoli, Italy\\
$^{3}$  \quad Miur: Ministero dell'Istruzione, dell'Università e della Ricerca, Italy 
}

\corres{Correspondence: valboz@sa.infn.it}

\abstract{We investigate binary lenses with $1/r^n$ potentials in the asymmetric case with two lenses with different indexes $n$ and $m$. These kinds of potentials have been widely used in several contexts, ranging from galaxies with halos described by different power laws to lensing by wormholes or exotic matter. In this paper, we present a complete atlas of critical curves and caustics for mixed binaries, starting from the equal-strength case, and then exploring unequal-strength systems. We also calculate the transitions between all different topology regimes. Finally we find some useful analytic approximations for the wide binary case and for the extreme unequal-strength case. }


\begin{document}
\section{Introduction}

Space-time is curved by the presence of massive bodies and this curvature influences the motion of the bodies themselves: this leads to a geometry in constant evolution. One of the consequences is that even light, supposed to be massless, bends its trajectory while passing close to a massive body. Einstein deduced it already in 1913, two years before his theory was completed \cite{Einstein1913}, and the British astronomer Arthur Eddington decided to exploit this intuition experimentally. On May 29th 1919 during a solar eclipse in Principe Islands he showed that stars moved from their position by the amount precisely predicted by general relativity. This great result was put in evidence by the main newspapers of that time, like \textit{Cosmic Time} that titled "Sun's gravity bends starlight" underlining the triumph of Einstein's theory. This was the first observation of \textit{gravitational lensing} \cite{Eddington1919}. 

Gravitational lensing is an important tool in astrophysics and in cosmology widely used to study both populations of compact objects (including exoplanets, black holes and other stellar remnants) \cite{Gaudi,Subaru}, and extended objects, such as galaxies, clusters of galaxies and large-scale structures \cite{Gavazzi2007,Hoekstra2008,Hoekstra2013,Kiblinger2015,Jeong2015}. Since most of the mysteries of our Universe do not show up in observations based on electromagnetic interactions, gravitational lensing is more and more employed to study the dark side of the Universe, including dark matter, dark energy, and any kind of exotic matter (such as wormholes) conjectured by theorists \cite{Schneider1992,Petters,Perlick,Schneider2006,Zak2010,Bozza2010}. 

Gravitational lensing effects by wormholes were investigated in Refs. \cite{Kim,Cramer}, with negative mass in Refs.\cite{Safonova,Eiroa,saf,saf2,Takahashi}
and with positive mass in Refs. \cite{Rahaman,Kuhfittig,Tejeiro,Nandi,Dey}. We want to remark that in 1973, Ellis and Bronnikov  independently found a massless wormhole (the Ellis wormhole) as a wormhole solution of the Einstein equations, see Refs. \cite{Ellis1973,Bronnikov1973}. 
Also spherically symmetric and static
traversable Morris-Thorne wormholes were analyzed in Refs. \cite{Morris,Thorne}. The most general extension
of the Morris-Thorne wormhole is the solution of the stationary and axially symmetric rotating Teo wormhole in Ref. \cite{Teo}, the first rotating wormhole solution, and this was the starting point for the investigation of gravitational lensing by rotating wormholes explored by Jusufi and Ovgun in Ref. \cite{Jusufi}. Tsukamoto and Harada studied the light rays passing through a wormhole in Ref. \cite{TsuHar}; Ohgami and Sakai studied the images of wormholes surrounded by optically thin dust in Ref. \cite{oghs1} in order to state if it is possible to identify wormholes by observing
shadows; this was also investigated in Ref. \cite{oghs2} in rotating dust flow.

The metric of the Ellis wormhole falls down asymptotically as $1/r^2$ and its deflection angle goes as the inverse square of the impact parameter $1/u^2$ as explored in Refs. \cite{Chetouani,perl,Abe,Bhattacharya,toki,thy,na,gib,Yoo}. Metrics falling as $1/r^n$ were investigated also by Kitamura et al. \cite{Kitamura} who found out that the deflection angle falls down with the same exponent as the metric: $\hat{\alpha}\sim1/u^n$ with $n>1$. Other investigations include Refs. \cite{Izumi,tkna,nia,th,kit,Bozza15}.
Power-law deflection terms can also be found in gravitational lensing in the presence of plasma \cite{biskog10,xinro17,xinro19,xinro19b,biskog20}.
Particular attention was posed on the study of caustics of $1/r^n$ binary lenses by Bozza and Melchiorre in Ref. \cite{Bozza_2016} and to the investigation of gravitational lensing by exotic lenses with a non-standard form of the equation of state or with a modified gravity theory by Asada \cite{Asada}. A new method of detecting Ellis wormholes by the use of the images of wormholes surrounded by optically thin dust was investigated by Ohgami and Sakai \cite{Ohgami}.

After the Event Horizon Telescope results \cite{Akiyama}, consisting in the detection of the shadow of a supermassive
black hole in the center of galaxy M87,  many authors tried to explore new frontiers, and an interesting new reference is from Tsukamoto and Kokubu \cite{Tsukamoto20}: they investigate the collision of two test particles in the Damour-Solodukhin wormhole spacetime where Damour and Solodukhin stated in Ref. \cite{Damour} that is not possible to distinguish black holes from wormholes with observations on a limited timescale.

From the side of binary galaxies as binary lenses we must cite the considerable work of Shin and Evans \cite{Shin} that discussed the critical curves and caustics in the case $n<1$. This applies to generic galactic halos and isothermal sphere in particular, as the limit $n\to0$.

Kovner investigated extremal solutions for a singular isothermal sphere with a tide (SIST) \cite{Kovner}; Evans and Wilkinson studied lens models for representing cusped galaxies and clusters, as isothermal cusps always generate a pseudocaustic \cite{EvansWilkinson}, while Rhie discussed pseudocaustics of various lens equations \cite{Rhie}. Wang and Turner studied strong gravitational lensing by spiral galaxies, modelling them as infinitely
thin uniform disks embedded in singular isothermal spheres \cite{WangTurner}, while Tessore and Metcalf investigated a general class of lenses following an elliptical power law profile \cite{Tes}. All these systems possess pseudocaustics that were also investigated by Lake and Zheng in gravitational lensing by a ring-like structure \cite{Lake}. Higher-order caustic singularities, such as the elliptic umbilic, were discussed by Aazami et al. \cite{Aazami}.

In this work we want to extend the symmetric structure already studied by Bozza and Melchiorre in Ref. \cite{Bozza_2016} for $1/r^n$ potentials, in which the two lenses have the same index $n$, to an asymmetric case in which the lenses have different indexes. This generalization is particularly useful in both scientific contexts described by $1/r^n$ potentials. In fact, we may have pairs of galaxies that have very different structures and thus different halo profiles, e.g. a dwarf galaxy as a satellite to a giant galaxy. On the other hand, if wormholes or other exotic objects exist, they might be part of a binary system with an ordinary star or other compact objects. The co-existence of objects with different $1/r^n$ potentials thus seems plausible in many situations, thus justifying the generalization we are going to undertake here. 

In Section 2 we give the lens equation for $1/r^n$ potentials for two exotic lenses with different $n$. In Section 3 we study critical curves and caustics presenting three main cases: equal-strength binary lenses,  unequal-strength binary and extreme unequal-strength binary lenses explaining the origin of the pseudocaustic and of the elliptic umbilic catastrophe for $mn<1$. In Section 4 we study the transitions between different caustic topologies. In section 5 we derive analytical approximations for the three cases analysed in Section 3 in order to have a deeper understanding in the caustic evolution, in its shape and size. Finally in Section 6 we draw our conclusions.

 \section{ Gravitational lensing by objects with $1/r^n$ potential}
 
Objects whose gravitational potential asymptotically falls as $1/r^n$ ($n\leq1$ for ordinary matter, $n>1$ for exotic matter) give rise to a deflection angle that goes as $\alpha\sim1/{|\theta|^n}$, where $\theta$ is the angular position at which the image is observed. The lens equation for a single lens, first studied by Kitamura et. al in Ref. \cite{Kitamura} and then generalized by Bozza and Postiglione in Ref. \cite{Bozza15}, is

\begin{equation}
\beta=\theta-\frac{\theta_E^{n+1}}{|\theta|^n}
Sign(\theta),
\end{equation}
where $\beta$ is the source angular position with respect to the center of the lens, $\theta_E$ is the Einstein radius of the lens, which depends on the specific parameters of the metric describing the object \cite{Kitamura} and the index $n$ is either the exponent of the halo profile for a normal matter distribution or the ratio between tangential and radial pressure, $n=-2p_t/p_r$, if we consider exotic matter  \cite{Bozza15}. 

We want to explore a system composed by two objects in the asymmetric case in which our lenses have different indexes, here indicated with $n$ and $m$. The binary lens equation is

\begin{equation}
   \Vec{\beta}=\Vec{\theta}-\theta_{E,A}^{n+1}\frac{\Vec{\theta}-\Vec{\theta_A}}{|\Vec{\theta}-\Vec{\theta_A}|^{n+1}}-\theta_{E,B}^{m+1}\frac{\Vec{\theta}-\Vec{\theta_B}}{|\Vec{\theta}-\Vec{\theta_B}|^{m+1}},
\end{equation}
where $\Vec{\theta_A}$ and $\Vec{\theta_B}$ are the coordinates of the two objects in the sky. 

We note that the Einstein radii ${\theta}_{E,A}$ and ${\theta}_{E,B}$, appear with different exponents for each lens. It is thus convenient to use ${\theta}_{E,A}$ as a unit of measure for angles and define the "strength ratio" as $\gamma=\theta_{E,B}/\theta_{E,A}$. We rewrite the lens equation as follows

\begin{equation}
   \Vec{\beta}=\Vec{\theta}-\frac{\Vec{\theta}-\Vec{\theta_A}}{|\Vec{\theta}-\Vec{\theta_A}|^{n+1}}-\gamma^{m+1}\frac{\Vec{\theta}-\Vec{\theta_B}}{|\Vec{\theta}-\Vec{\theta_B}|^{m+1}}.\label{general}
\end{equation}

Now we introduce complex coordinates \cite{Witt1990}
\begin{equation}
    \zeta= \beta_1  + i\beta_2; z= \theta_1  + i\theta_2
\end{equation}

We take the mid-point between the two lenses as the origin of the coordinates, and orient the real axis along the line joining the two lenses. We thus set $z_A=-s/2$ and $z_B=s/2$, where $s$ is the normalized angular separation between the lenses. The lens equation becomes 
    
\begin{equation}
 \zeta=z-\frac{1}{\left({z}+\frac{s}{2}\right)^{\frac{n-1}{2}} \left(\bar{z}+\frac{s}{2}\right)^{\frac{n+1}{2}}}-\frac{\gamma^{m+1}}{\left(z-\frac{s}{2}\right)^{\frac{m-1}{2}} \left(\bar{z}-\frac{s}{2}\right)^{\frac{m+1}{2}}}\label{lenses}
\end{equation}

The Jacobian determinant of the lens map in complex notation is given by

\begin{equation}
    J(z,\bar{z})=\left|\frac{\partial\zeta}{\partial z}\right|^2-\left|\frac{\partial\zeta}{\partial \Bar{z}}\right|^2,\label{lenseq}
\end{equation}
which in our case becomes

\begin{equation}
  \begin{split}
J &=\Bigg[1+\frac{1}{2}\left( \frac{n-1}{\left(z+\frac{s}{2}\right)^{\frac{n+1}{2}} \left(\bar{z}+\frac{s}{2}\right)^{\frac{n+1}{2}}} +\frac{\gamma^{m+1}(m-1) }{\left(z-\frac{s}{2}\right)^{\frac{m+1}{2}} \left(\bar{z}-\frac{s}{2}\right)^{\frac{m+1}{2}}}\right)\Bigg]^2\\  
&-\frac {1}{4}\left|\frac{n+1}{\left(z+\frac{s}{2}\right)^{\frac {n+3} {2}}\left(\bar{z}+\frac{s}{2}\right)^{\frac{n-1}{2}}}+\frac{\gamma^{m+1}(m+1) }{\left(z-\frac{s}{2}\right)^{\frac{m+3}{2}} \left(\bar{z}-\frac{s}{2}\right)^{\frac{m-1}{2}}}\right|^2\\
\end{split}.
\end{equation}

We note that the structure of the Jacobian becomes more complicated with respect to the ordinary point-lenses ($m=n=1$), in which many terms disappear. We thus expect a correspondingly richer phenomenology. The Schwarzschild case was already explored by Schneider and Weiss in Ref. \cite{Schneider} for lenses with the same mass and by Erdl and Schneider for lenses with different masses \cite{Erdl}; Bozza and Melchiorre investigated the case $m=n$. In order to compare our results with theirs, it is important to note that there is no notion of a combined total Einstein radius when the two lenses have different indexes for their potentials. Therefore, the notation introduced there, with $\epsilon_{i}$ as the ratio of the individual lens strength to the total strength cannot be replicated here. Their results were expressed in terms of the ratio $q=\epsilon_{B}/\epsilon_{A}$. The relation between our parameter $\gamma=\theta_{E,B}/\theta_{E,A}$ and $q$ is just $\gamma^{m+1}=q$. As a practical example, the Einstein radius scales as $\sqrt{q}$ in the Schwarzschild case, where $q$ becomes the mass ratio of the two lenses.
  
\section{Critical curves and caustics}
  
The condition $J(z)=0$ defines the critical curves on the lens plane. By applying the lens map on critical points we find the corresponding points on the source plane, which form the caustics. Critical curves and caustics are of fundamental importance to understand how gravitational lensing works. When a source crosses a caustic, a new pair of images is created on the corresponding point in the critical curve. Therefore, caustics bound regions with a different number of images. Critical curves distinguish regions in which images have opposite parities. 

Our model contains four parameters: the indexes of the two potentials $n$, $m$, the separation between the two lenses $s$, and the ratio of the two Einstein radii $\gamma$. In order to start the exploration of this parameter space, we first  analyze the equal-strength case with $\gamma=1$, and then move to unequal strength cases. 
 
In all plots presented in this paper, we keep $n=1$ fixed for the first lens (ordinary Schwarzschild lens), with variable $m$ for the second lens: $m=0,0.5,1,2,3$ (we remind that $m=0$ is the singular isothermal sphere, already investigated by Shin and Evans in Ref. \cite{Shin}, galactic halos are in the range $0<m<1$ and $m=2$ corresponds to the Ellis wormhole; objects with $m>1$ require exotic matter).
 
Critical curves are obtained by the contour plot of the Jacobian determinant and these contours are then mapped through the lens equation in order to get the caustics. All computations are performed by \textit{Wolfram Mathematica 11} \footnote{https://www.wolfram.com/mathematica/}.
 
\subsection{Equal-strength binaries}

In the equal-strength case, we set $\gamma=1$, which means that $\theta_{E,A}=\theta_{E,B}$: both lenses would generate a critical curve with the same radius if they were isolated.

For the standard binary Schwarzschild lens \cite{Schneider}, we know that three topologies exist:
\begin{itemize}
    \item [-] close separation, for $s<s_{CI}$;
    \item [-] intermediate separation, for $s_{CI}<s<s_{IW}$;
    \item [-] wide separation, for $s>s_{IW}$;
\end{itemize}
and the two transitions are $s_{CI}=1$ and $s_{IW}=2\sqrt{2}$ in our units.

We find that these three topologies persist for any values of $n$ and $m$, although the boundary values may vary somewhat. In order to illustrate the evolution of critical curves and caustics in intelligible figures, we present the plots for different values of $m$ at fixed values of separation $s$, starting from wide separation binaries and then moving the two lens closer.

\begin{figure}[H]
\centering
\includegraphics[height=6.2 cm]{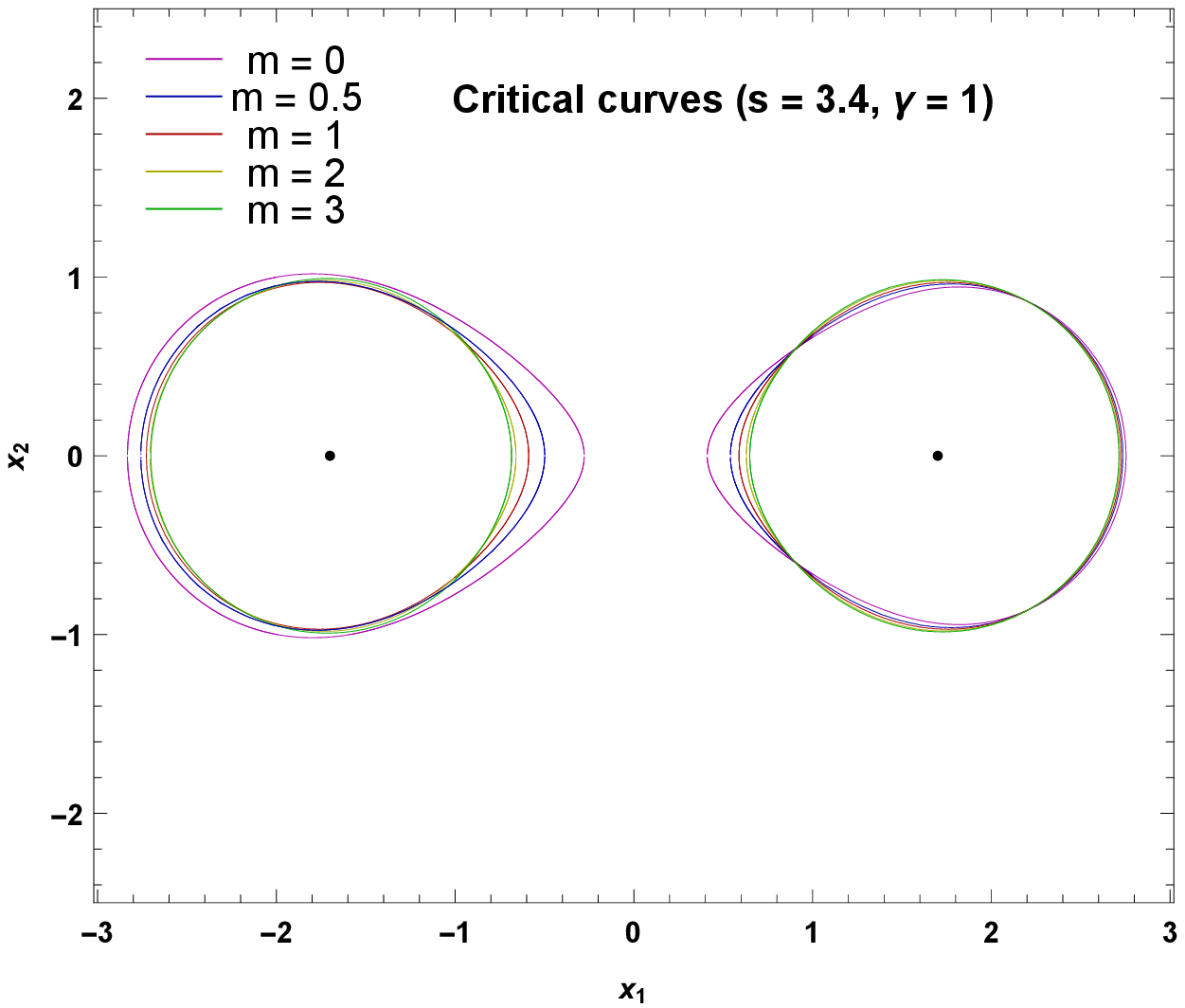}
\includegraphics[height=6.2 cm]{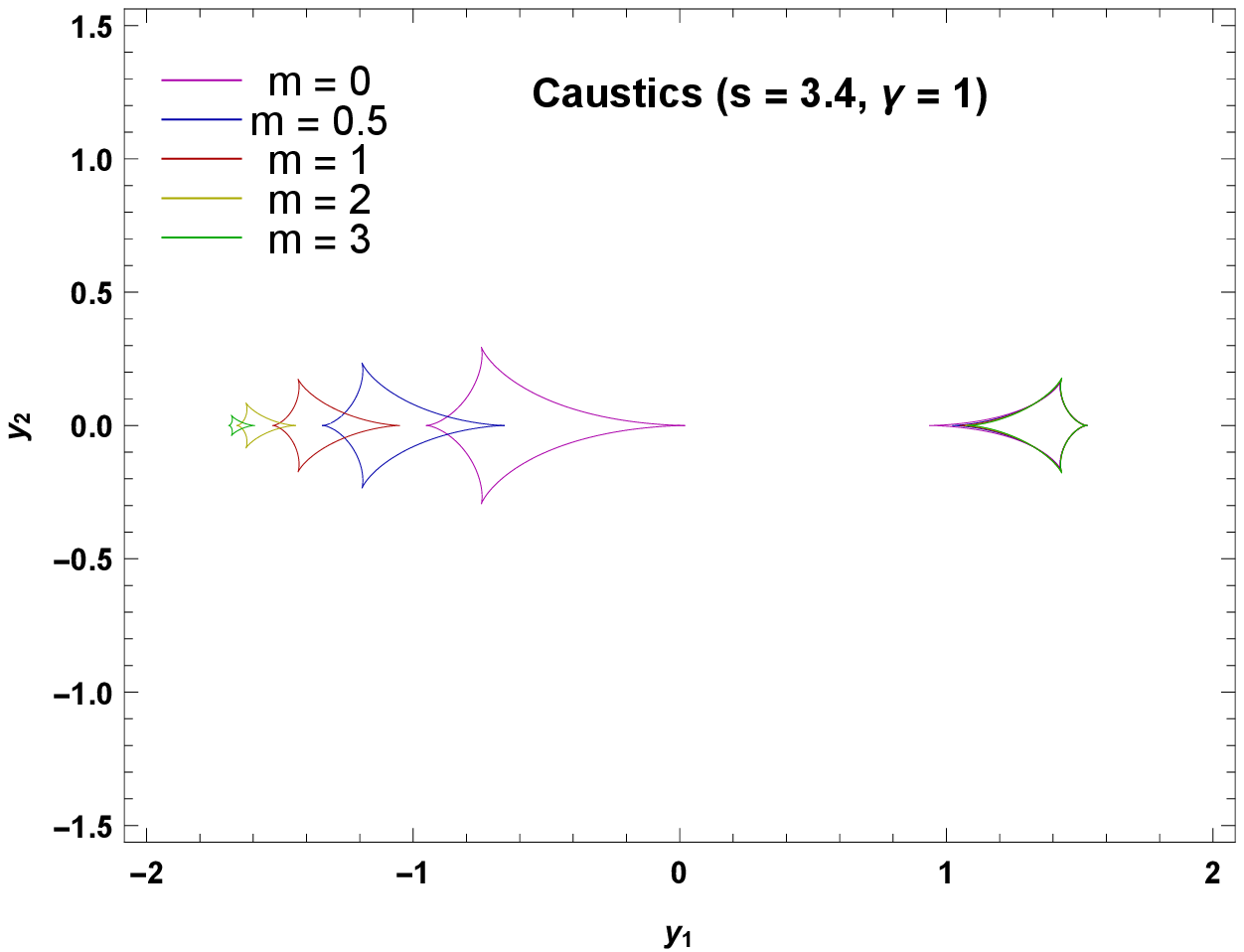}
\caption{Critical curves and caustics in the equal-strength binary, wide separation. Here and in the following figures the lens on the left side has $n = 1$ and the lens on the right side has variable $m$, coherently with Eq. (\ref{lenses}).}\label{wi}
\end{figure}

First, in Fig. \ref{wi} we have two lenses at wide separations for $s=3.4$. Here we clearly see how the Einstein ring of each lens is distorted by the presence of the partner lens. Comparing the critical curves obtained at different values of the index $m$, we clearly see that the distortion is stronger for small values of $m$. This is a direct consequence of the fact that the potential decays more steeply for larger $m$ and thus the first lens feels a weaker tidal field from the second lens. This is particularly evident for the caustic of the first lens, which becomes very small at $m=3$, while it becomes larger and more shifted at $m=0$. The caustic of the second lens is almost independent of $m$. In practice, the shape and the size of the caustic is mostly determined by the tidal field of the first lens, which we are keeping fixed with $n=1$.

\begin{figure}[H]
\centering
\includegraphics[height=6 cm]{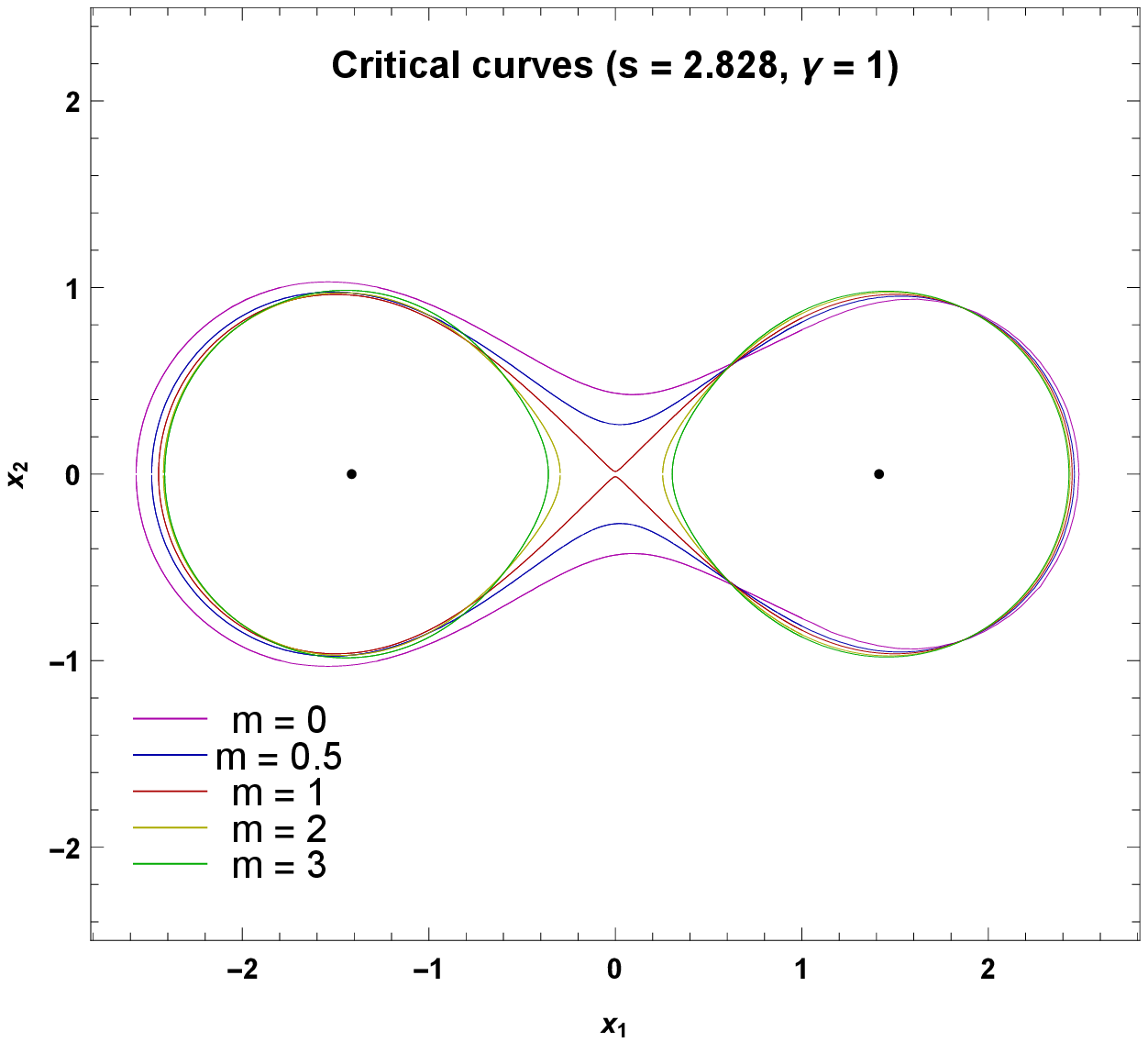}
\includegraphics[height=6 cm]{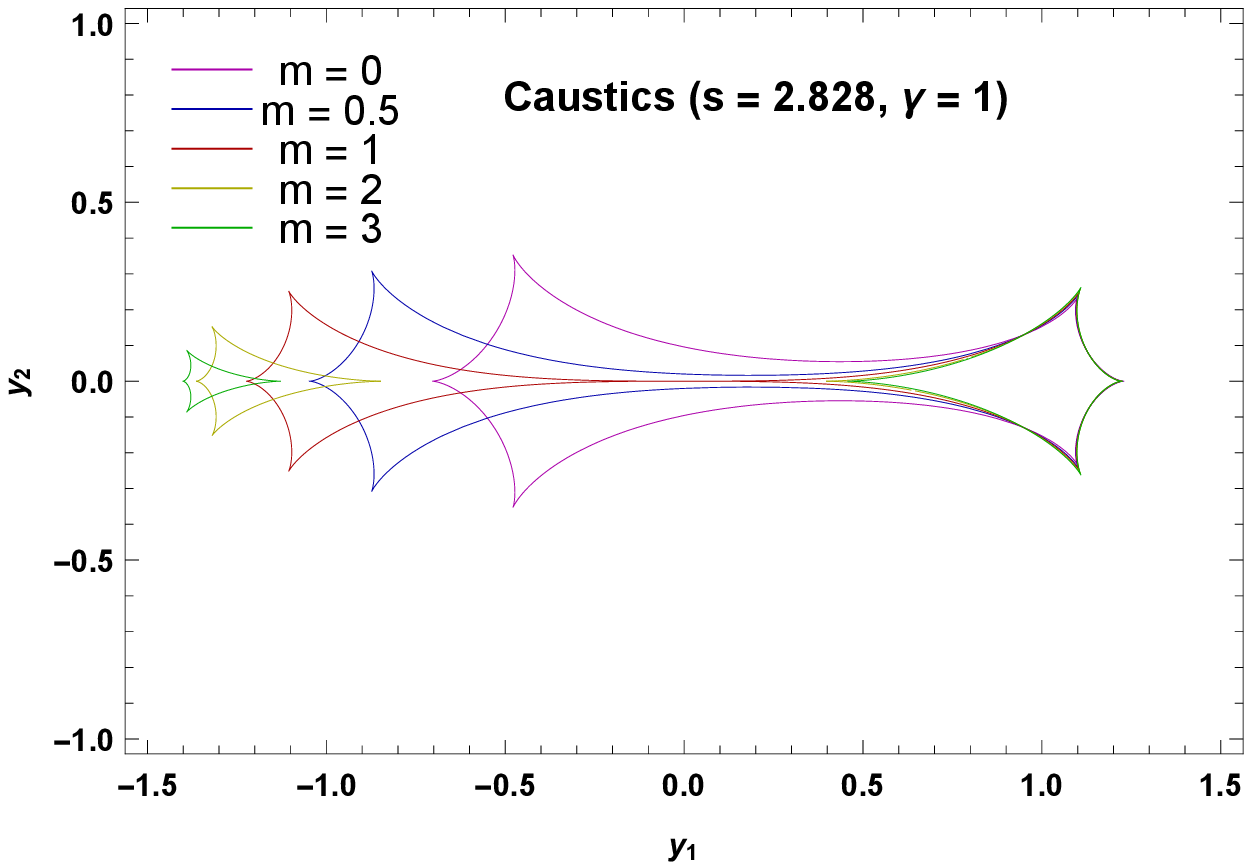}
\caption{Critical curves and caustics in the equal-strength binary, intermediate-wide transition.}\label{intwide}
\end{figure}

In Fig. \ref{intwide} we show critical curves and caustics for $s=2\sqrt{2}$, which corresponds to the intermediate-wide transition in the standard $n=m=1$ case. In fact, the red curves show the typical beak-to-beak singularity in the origin.
For $m<1$ we are already in the intermediate regime, while for $m>1$ we are still in the wide regime. As explained before, the fact that the intermediate regime extends to larger separations for $m<1$ is a consequence of the slower decay of the potential.  

\begin{figure}[H]
\centering
\includegraphics[height=6.7 cm]{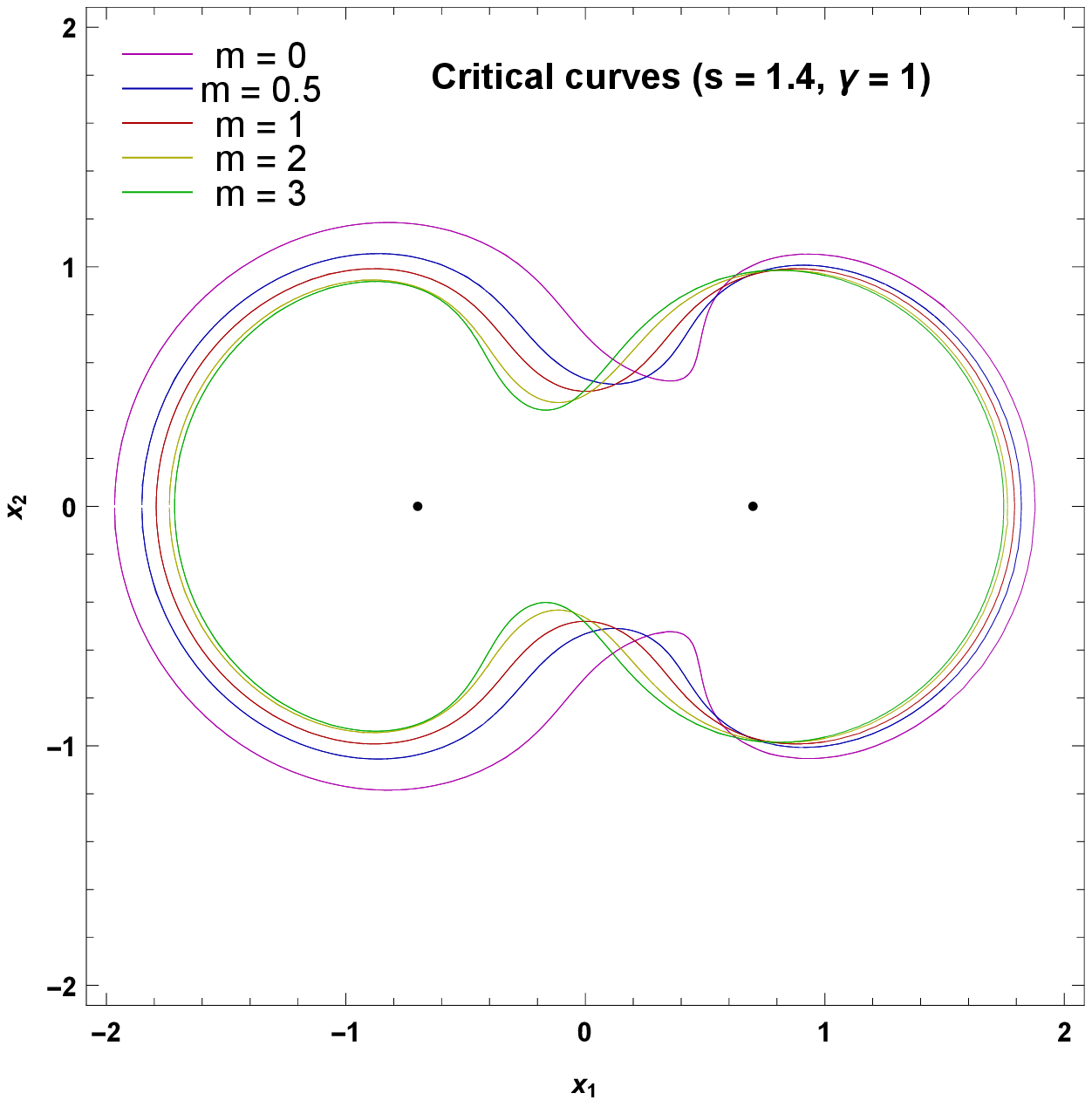}
\includegraphics[height=6.7 cm]{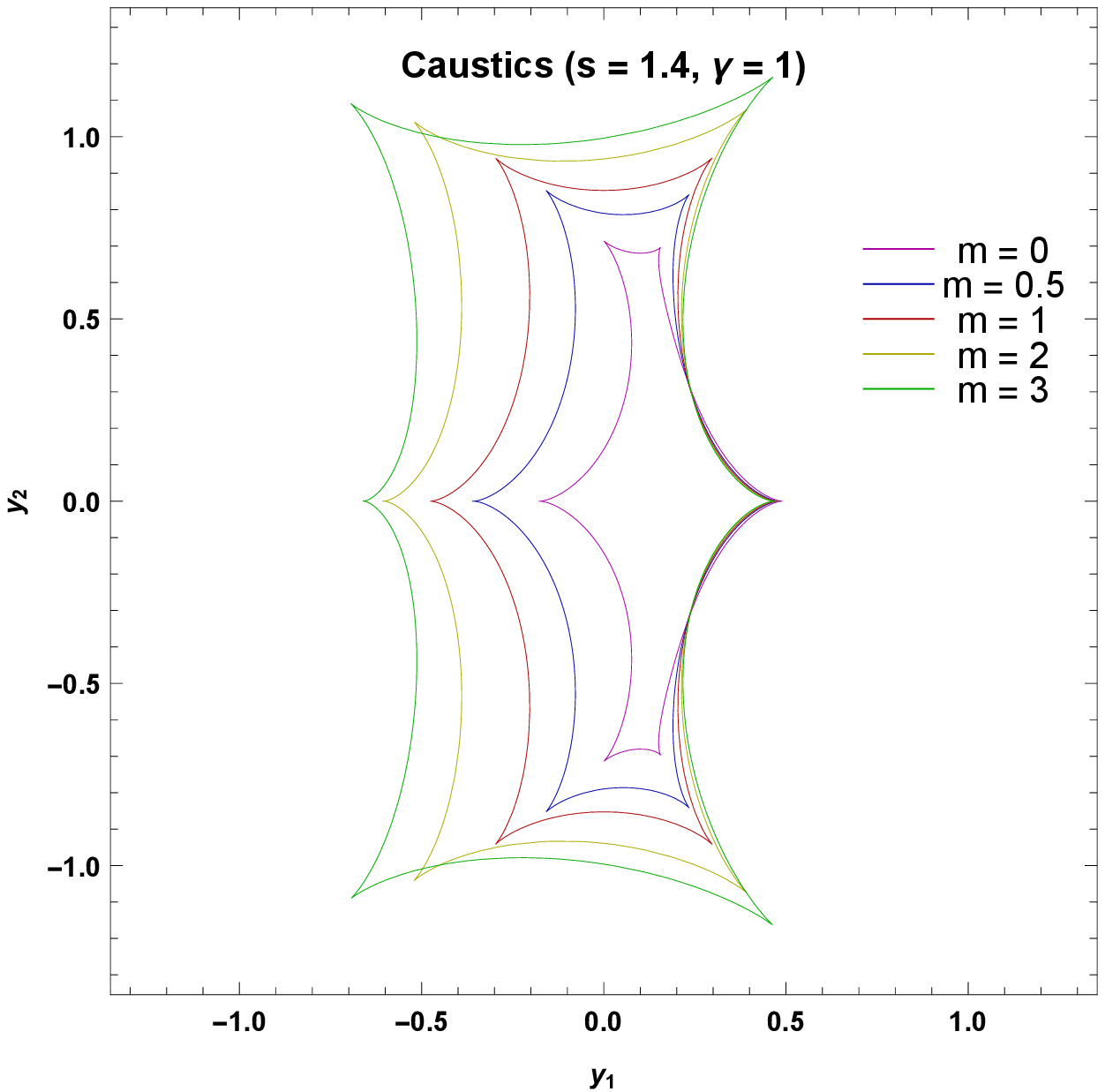}
\caption{Critical curves and caustics in the equal-strength binary, intermediate separation.}\label{int}
\end{figure}

In Fig. \ref{int} we see an intermediate separation at $s=1.4$: critical curves are larger for smaller values of $m$, while caustics are larger for increasing $m$. Indeed, we are starting to see some kind of inversion in the behavior of the lenses. Steeper profiles are going to dominate at smaller separations, as will be more evident in the incoming figures.

\begin{figure}[H]
\centering
\includegraphics[height=6.7 cm]{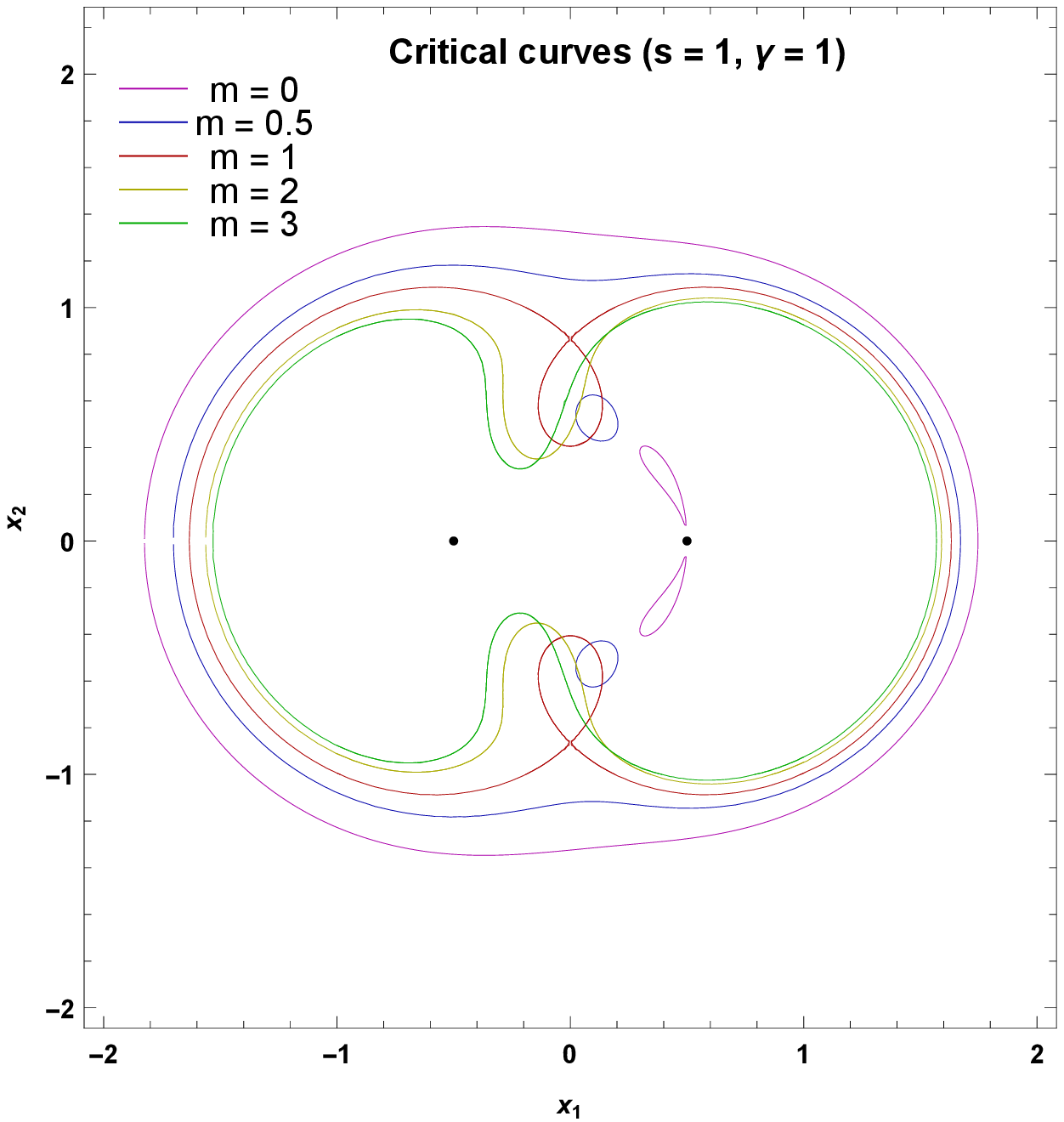}
\includegraphics[height=6.7 cm]{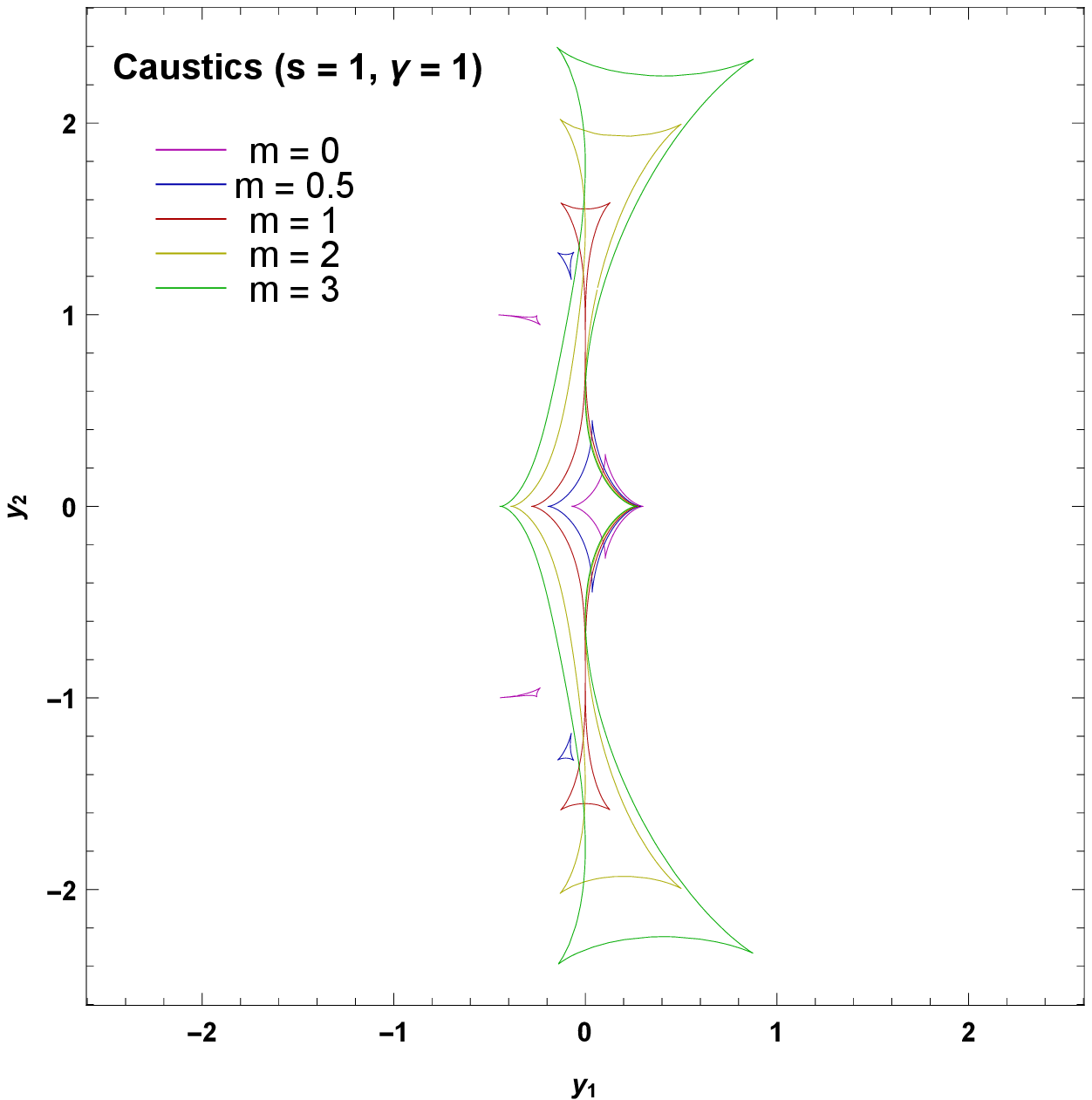}
\caption{Critical curves and caustics in the equal-strength binary, close-intermediate transition.}\label{gamma1ci}
\end{figure}

In Fig. \ref{gamma1ci} we show the critical curves and caustics at $s=1$, which corresponds to the close-intermediate transition for the standard $n=m=1$ case. In fact, the red curve shows the two symmetric beak-to-beak singularities. Contrary to the previous transition, now the $m<1$ caustics are already in the close regime, with small oval critical curves generating small triangular caustics. The $m>1$ curves are still in the intermediate regime. Following the same reasoning, $m<1$ lenses become subdominant in this regime and their influence on the whole system is smaller. In this regime, we also find the \textit{elliptic umbilic catastrophe} that we shall discuss in Subsection \ref{eo}.
 
\begin{figure}[H]
\centering
\includegraphics[height=6.7 cm]{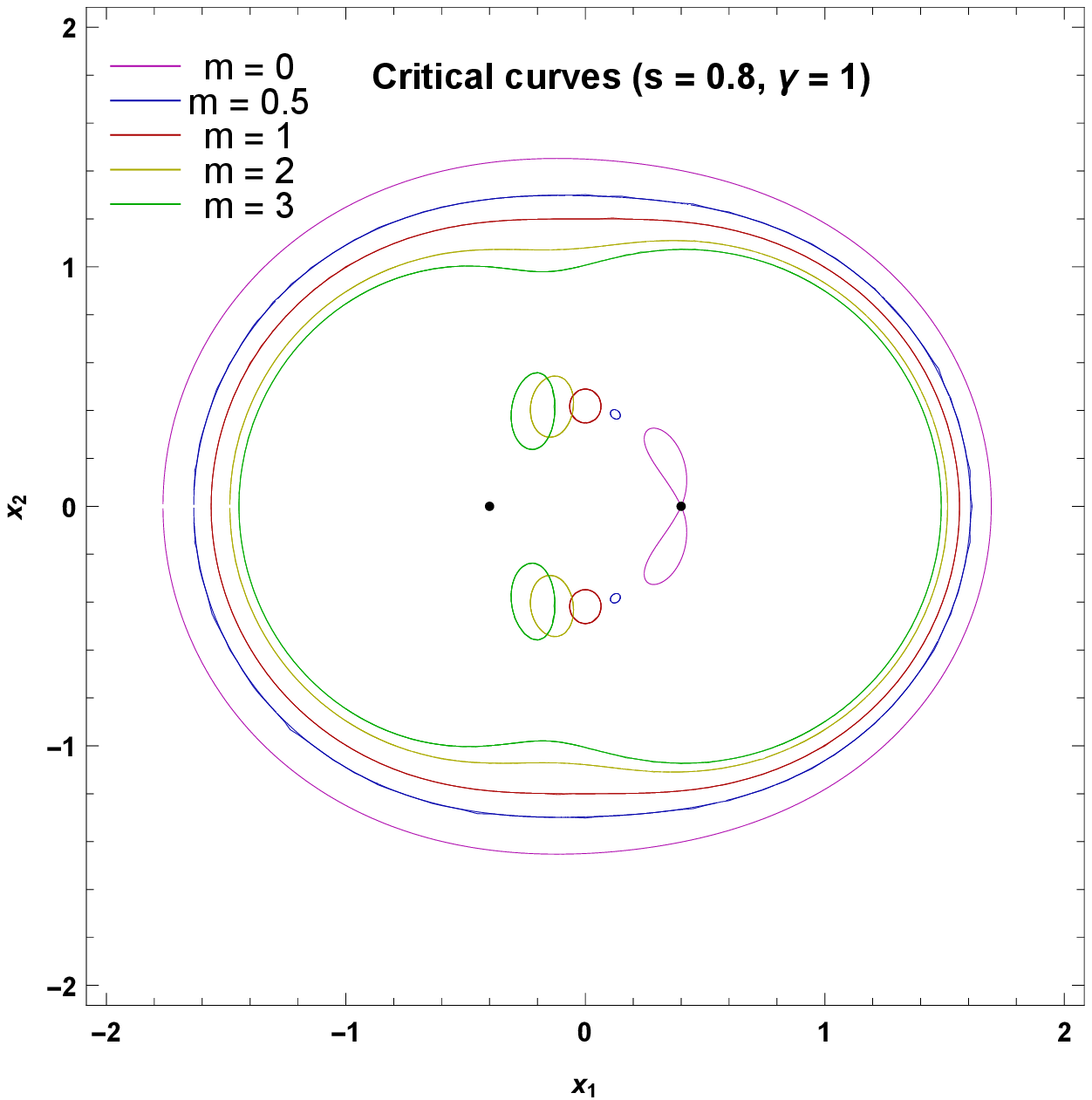}
\includegraphics[height=6.7 cm]{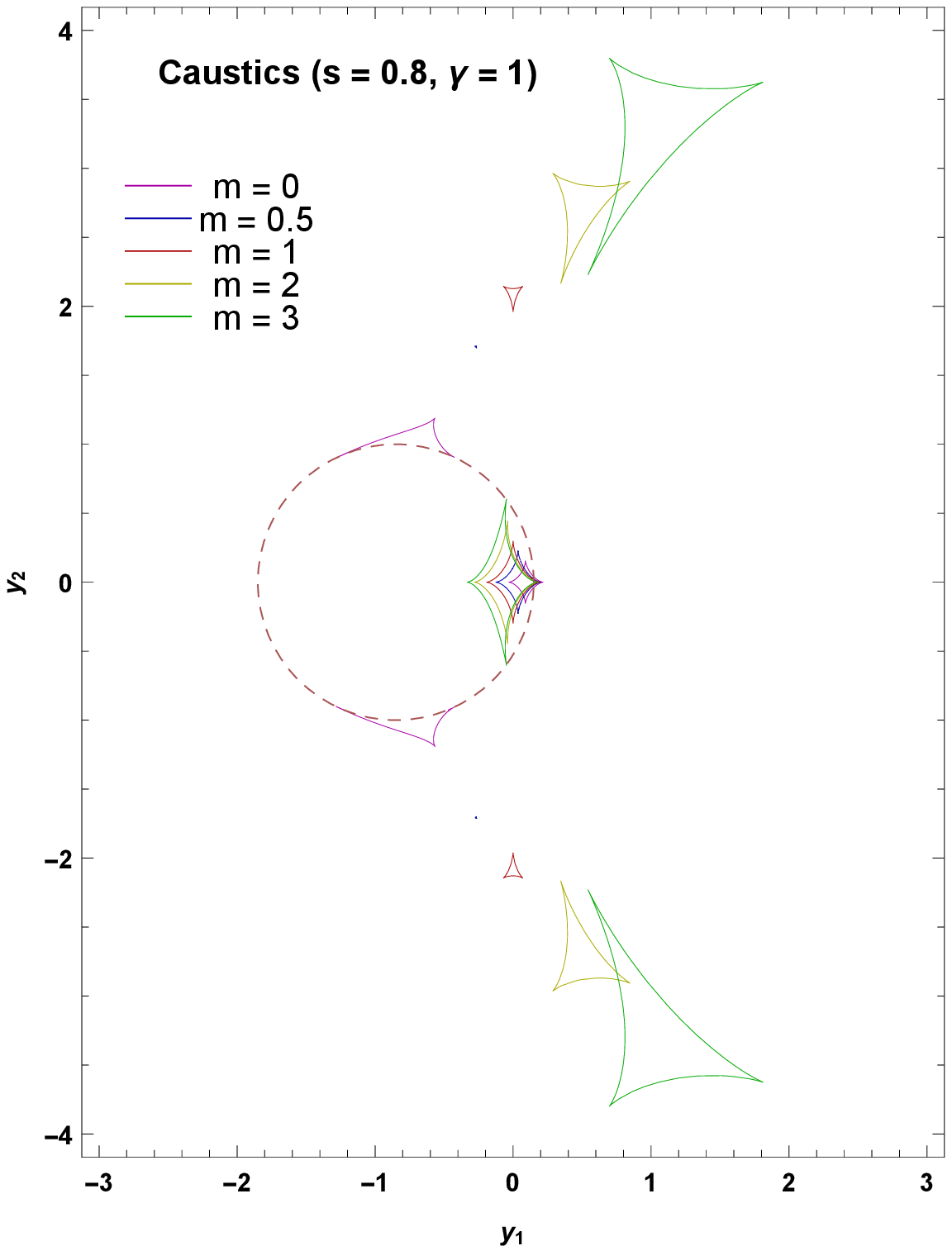}
\caption{Critical curves and caustics in the equal-strength binary, close separation. Dashed magenta circle indicates the pseudocaustic for $m=0$.}\label{gamma1close}
\end{figure}

In Fig.\ref{gamma1close} we can see critical curves and caustics in the close separation, for $s=0.8$.

Primary critical curves are big ovals that become smaller as $m$ increases. In fact, $m>1$ curves tend to be closer to an intermediate regime. Secondary critical curves are small ovals  that move far from the second lens, in the left direction, for $m>1$; for $m=0$ (magenta line) they converge on the second lens with the shape of a lemniscate: in this point the lens map is indeterminate and the corresponding caustics remain open on a circle that is called pseudocaustic. We shall discuss this structure in subsection \ref{pseudo}.

On the other side we find that the central caustics have the typical 4-cusps shape and they become smaller as $m$ decreases. Secondary caustics are always triangular but are considerably larger for $m>1$, a fact that was already stressed in Ref. \cite{Bozza_2016}. Note that for $m>1$ triangular caustics move right, while the central caustic is slightly displaced to the left. The opposite occurs for $m<1$. We can find a similar  behavior for standard Schwarzschild binaries with unequal masses. In practice, although we started with the same Einstein radius for both lenses, steeper profiles ($m>1$) behave similarly to heavier masses in this regime, while shallower profiles ($m<1$) behave as lighter masses.

\subsubsection{The Pseudocaustic\label{pseudo}}

A pseudocaustic is a closed curve on the source plane that exists for singular distributions with zero core radius. In the singular limit, the radial critical curve collapses onto the center of the lens, leaving no space for the dim central type III image. When the source crosses the corresponding radial caustic, only one more image forms, while the other image is degenerate with the center of the lens. The radial caustic is then named pseudocaustic, since it behaves differently from normal caustics \cite{Kovner,EvansWilkinson,Rhie,WangTurner,Tes,Lake}.

In the binary case, a pseudocaustic may still exist in the singular limit $m=0$. Through an analytical exploration we find out the points where the two secondary triangular caustics touch the pseudocaustic. 

The pseudocaustic is generated by critical curves collapsing to the center of the lens when $m=0$. In order to explore what happens around the center of the second lens, we set

\begin{equation}
    z=\frac{s}{2}+\text{$\epsilon_1$}+i \text{$\epsilon_2$};
\end{equation}
we expand around zero at $1/\epsilon$ order and then we solve with respect to $\epsilon_2$. 

We get two symmetric solutions

\begin{equation}
   \text{$\epsilon_2$}= \pm\frac{\sqrt{s^2+1} }{\sqrt{1-s^2}}\text{$\epsilon_1$}
\end{equation}

These solutions are two straight lines that cross at the origin of the system, and their angular coefficient is real only for $s<1$. This means that the two small oval critical curves will touch the center of the $m=0$ lens for separations in this regime. By substituting in the lens equation we find

\begin{equation}
\zeta=\frac{s}{2}-\frac{1}{s} \pm\gamma\sqrt{-s^2\pm\sqrt{s^4-1}}\label{pseudoc}
\end{equation}

These are the coordinates of the four contact points of the two triangular caustics with the pseudocaustic of radius $\gamma$ and center $(\frac{s}{2}-\frac{1}{s},0)$. The term $\frac{1}{s}$ shifts the caustic to the left side with respect to the position of the second lens  $\frac{s}{2}$. 

If the source only crosses the pseudocaustic, we have the sudden creation of one image of negative parity; if the source crosses a triangular caustic first and then the pseudocaustic, we have the formation of two images and the one inside the lemniscate (with positive parity) collapses on the lens.

\subsubsection{The Elliptic Umbilic}\label{eo}

As shown in Refs. \cite{Shin,Bozza_2016}, in the range $0\leq m<1$ an \textit{elliptic umbilic catastrophe} exists in the close separation. In an elliptic umbilic, the size of the small oval critical curves goes to zero and then grows up to finite size again. The catastrophe lies on a circle centered in the origin of the system, at the mid-point between the two lenses, and passing through them. It occurs at a specific separation $s$, which depends on the other parameters of the lens $\gamma$, $m$, $n$.

To find out the separation $s$, for any $m$ and $n$, at which the catastrophe occurs we proceed as follow: first we write the system of equations 
\begin{equation}
\begin{sistema}
$J=0$ \\
$$\frac{\partial J}{\partial z}=0$$
\end{sistema}\label{sys}
\end{equation}
along the circle, i.e. we set
\begin{equation}
    z=s\frac{e^{i\theta}}{2}.
\end{equation}

Then we introduce a new angular variable $t$ in order to simplify our computation
\begin{equation}
    t=\frac{\sin ^{m+1}(\theta/2)}{\cos ^{n+1}(\theta/2)}
\end{equation}

From Eq. (\ref{sys}), we get the angular position of the elliptic umbilic
 
\begin{equation}
     t=\frac{(m+1) \gamma ^{m+1}}{(n+1) s^{n-m}}
\end{equation}

and then we finally obtain the value of $s$ at which the catastrophe happens

\begin{equation}
  s_{euc}= \left(\frac{1-m n}{m+1}\right)^{\frac{1}{n+1}} \sqrt{1+\frac{\gamma ^2 (m+1)^{\frac{2}{n+1}}}{(n+1)^{\frac{2}{m+1}} (1-m n)^{\frac{2 (m-n)}{(m+1)(n+1)}}}}
\end{equation}

\begin{figure}[H]
\centering
\includegraphics[height=6.7 cm]{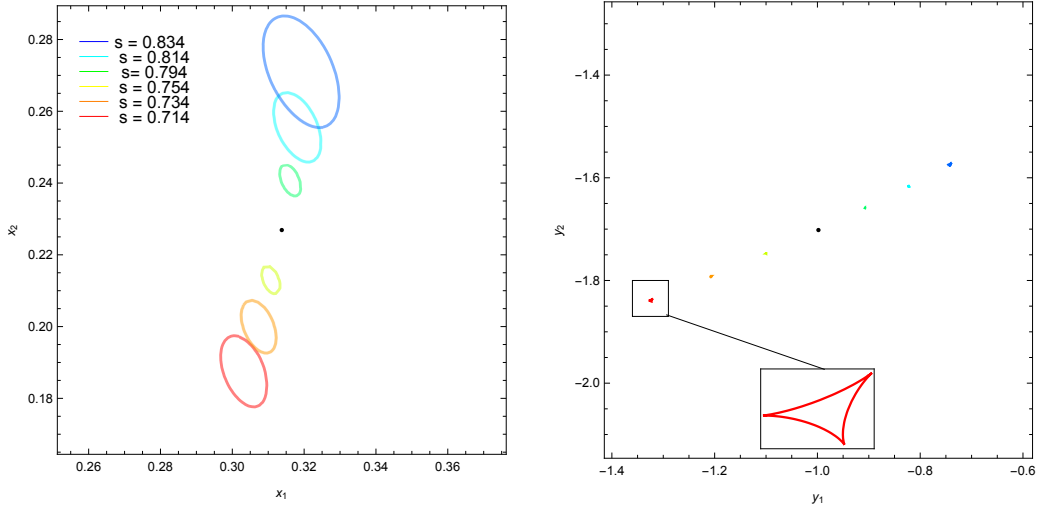}
\caption{The elliptic umbilic catastrophe for $n=2$, $m=0.25$ and $0.714\le s \leq 0.834$ in steps of $0.02$, one colour for each $s$ from red to blue. The separation at which the catastrophe occurs is $s_{euc}=0.774$. Critical curves on the left side panel, caustics on the right side panel.}\label{eu}
\end{figure}

Note that the solution exists for $mn<1$. In mixed binaries, we may have an elliptic umbilic also when one of the two lenses has a steep potential with $n>1$. In order to illustrate this, we choose $n=2$ (exotic matter) and $m=0.25$ (a possible galactic halo). We can see, in Fig. \ref{eu}, a zoom on the small oval critical curve for $0.714\le s \leq 0.834$ in steps of $0.02$. The separation at which the catastrophe occurs is $s_{euc}=0.774$. The critical curve shrinks to zero size for growing $s$, from $s=0.714$ to $s=0.774$ (lower curves) and then it grows up again. The corresponding triangular caustics behave similarly. 
 
\subsection{Unequal-strength binary\label{uneq}}

In the unequal-strength binary case, in order to keep contact with the previous work, we consider $q=0.1$, as in Ref. \cite{Bozza_2016}, so in terms of the ratio of the Einstein radii, our strength ratio is $\gamma=\sqrt{0.1}$. We need to multiply $s$ in Ref. \cite{Bozza_2016} by a factor$\sqrt{q+1}=\sqrt{1.1}$, so the transitions between different topologies for $n=m=1$ occur as follows:
\begin{itemize}
    \item [-] close-intermediate transition, $s_{CI}=0.807$;
    \item [-] intermediate-wide transition, $s_{IW}=1.772$.
\end{itemize}

Therefore, in this subsection we have the standard lens on the left with bigger Einstein radius than the lens on the right, for which we vary the potential index $m$.

We shall discuss each value of $m$ in the range $0\leq m\leq 3$ in detail.

\begin{figure}[H]
\centering
\includegraphics[height=6.7 cm]{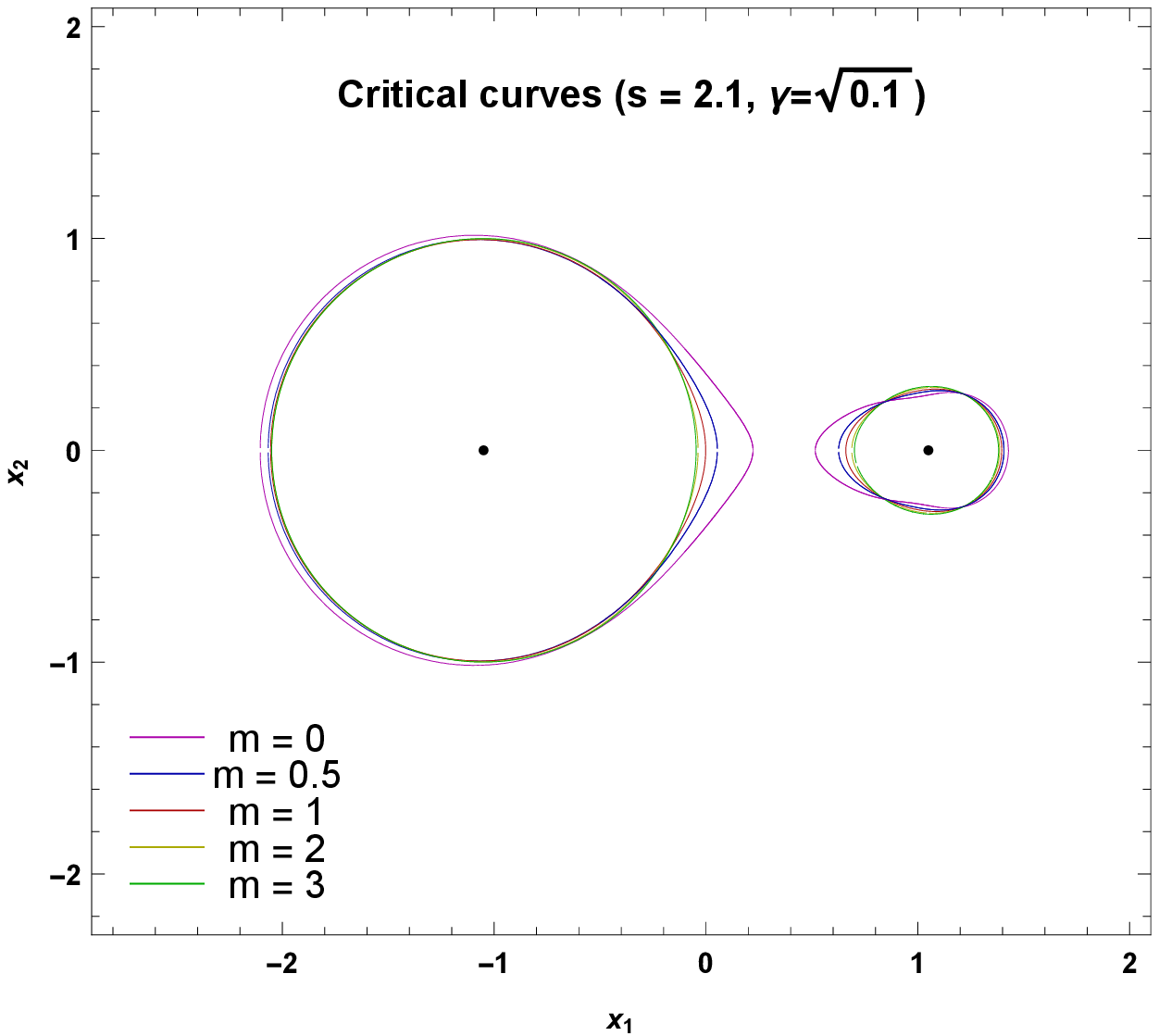}
\includegraphics[height=6.7 cm]{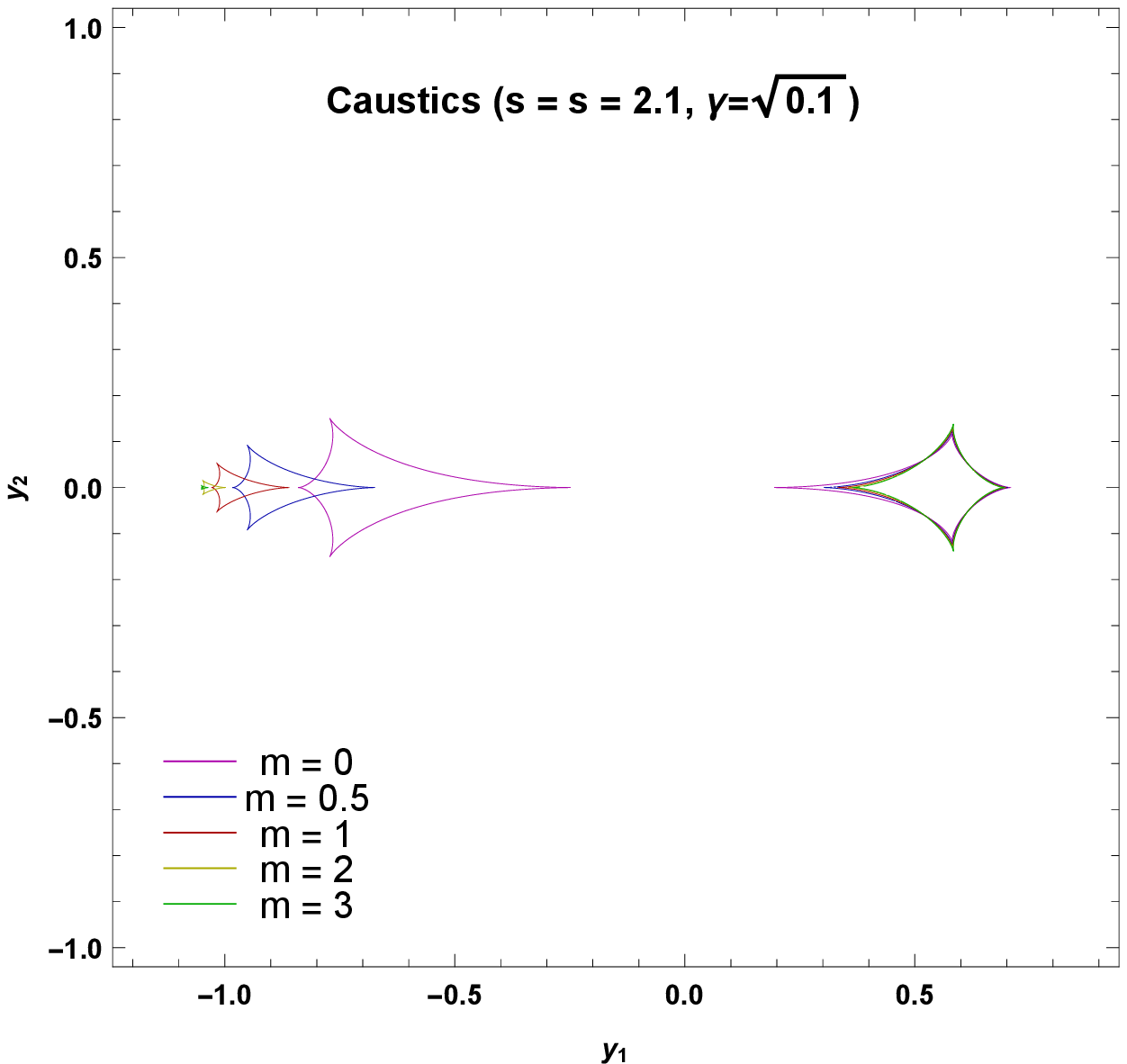}
\caption{Critical curves and caustics in the unequal-strength binary, wide separation. Here and in the following figures the lens on the left has $n = 1$ and the lens on the right has variable $m$.}\label{wide01}
\end{figure}

In Fig. \ref{wide01} we show the wide separation for $s=2.1$. Critical curves are separated and slightly deformed. The caustic of the left side lens is smaller since the tidal field from the right side lens is normally weaker. However, for $m<1$ the potential decays slower enough to make the left side caustic bigger than the caustic on the right.

\begin{figure}[H]
\centering
\includegraphics[height=6.7 cm]{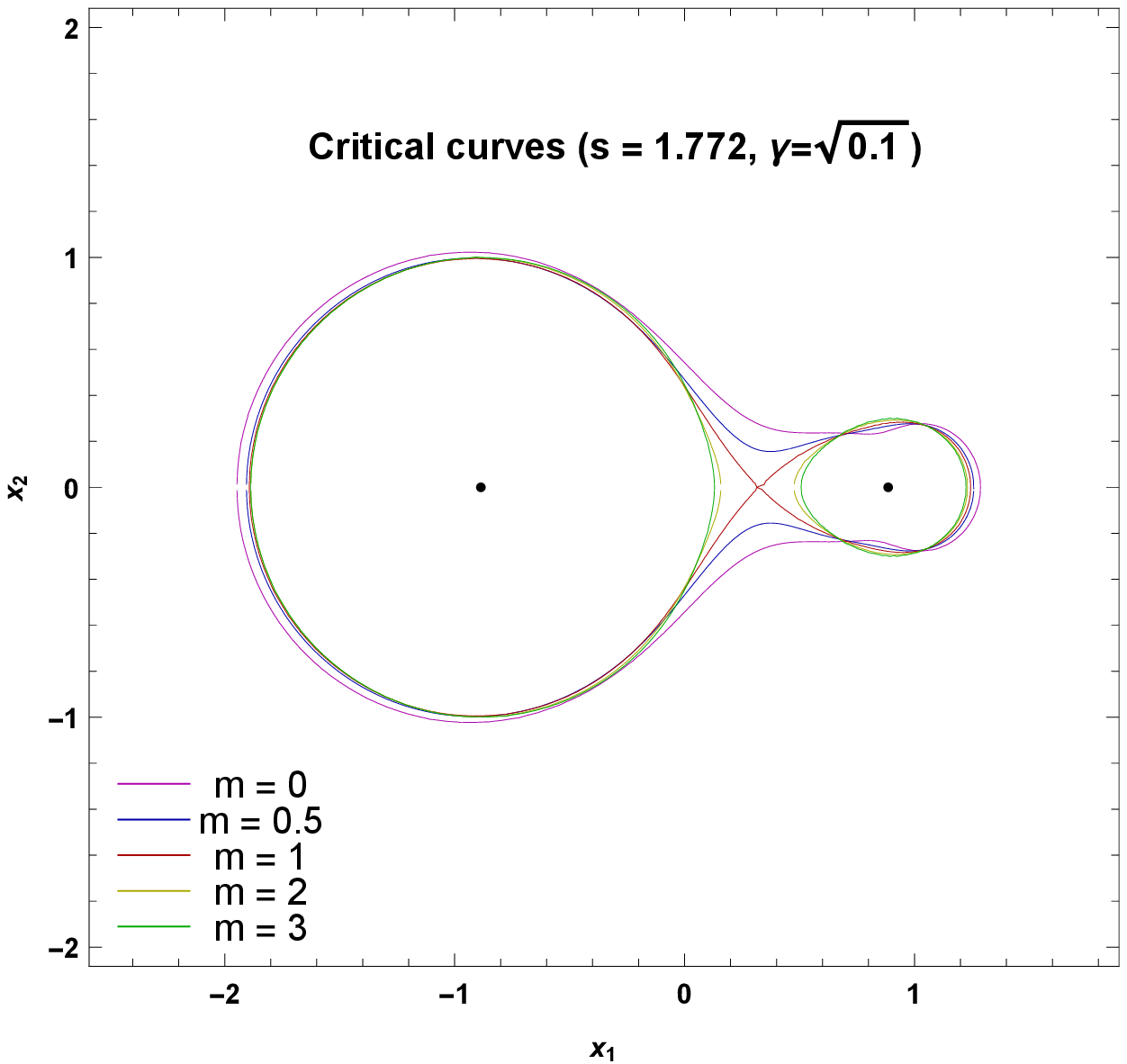}
\includegraphics[height=6.7 cm]{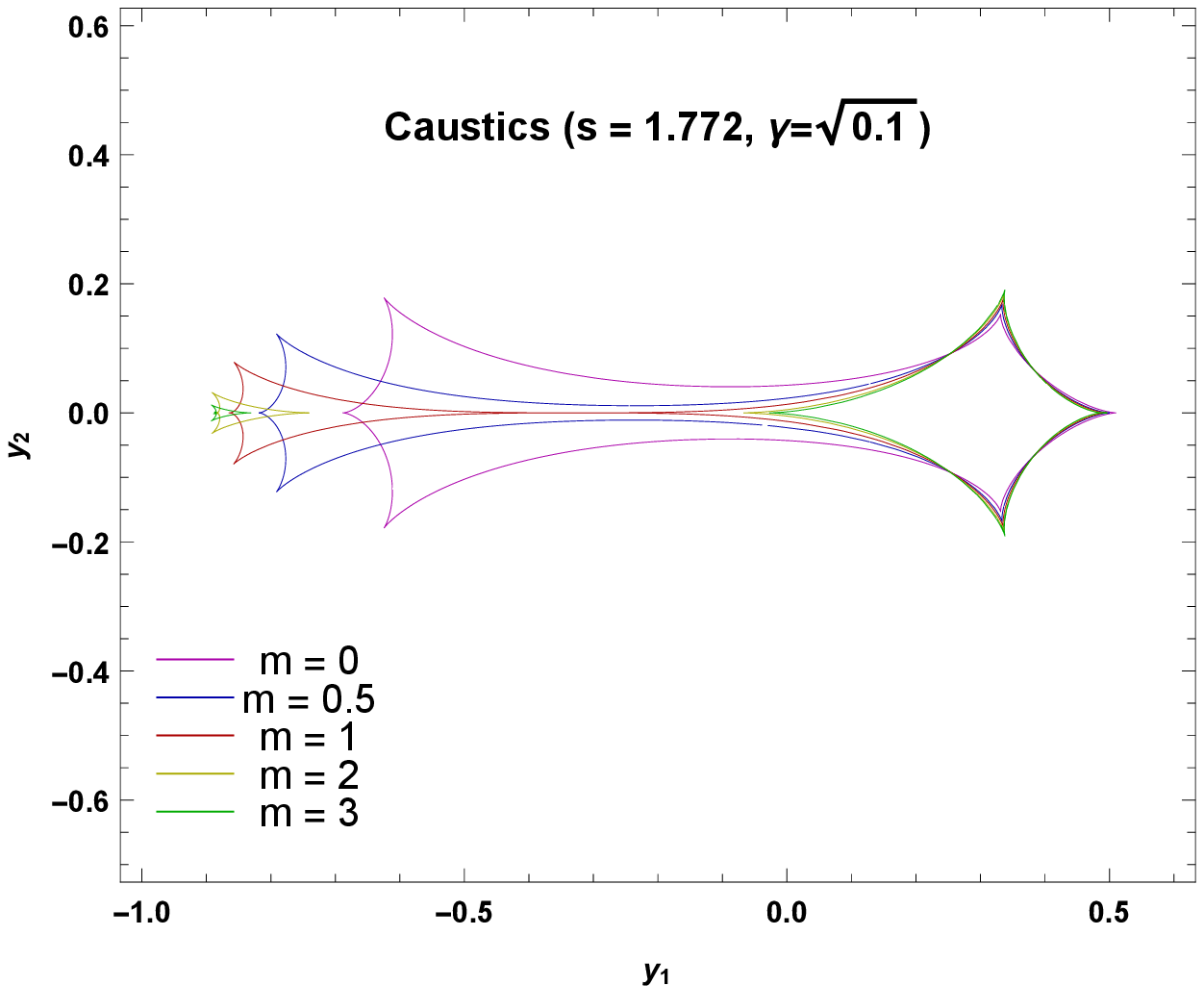}
\caption{Critical curves and caustics in the unequal-strength binary, intermediate-wide transition.}\label{iw01}
\end{figure}

In Fig. \ref{iw01} we can see the intermediate-wide transition at $s=1.772$. For $m<1$ (magenta and blue lines) the transition to the intermediate regime has already occurred, while we are still in the wide regime for $m>1$ (green and yellow lines).

\begin{figure}[H]
\centering
\includegraphics[height=6.7 cm]{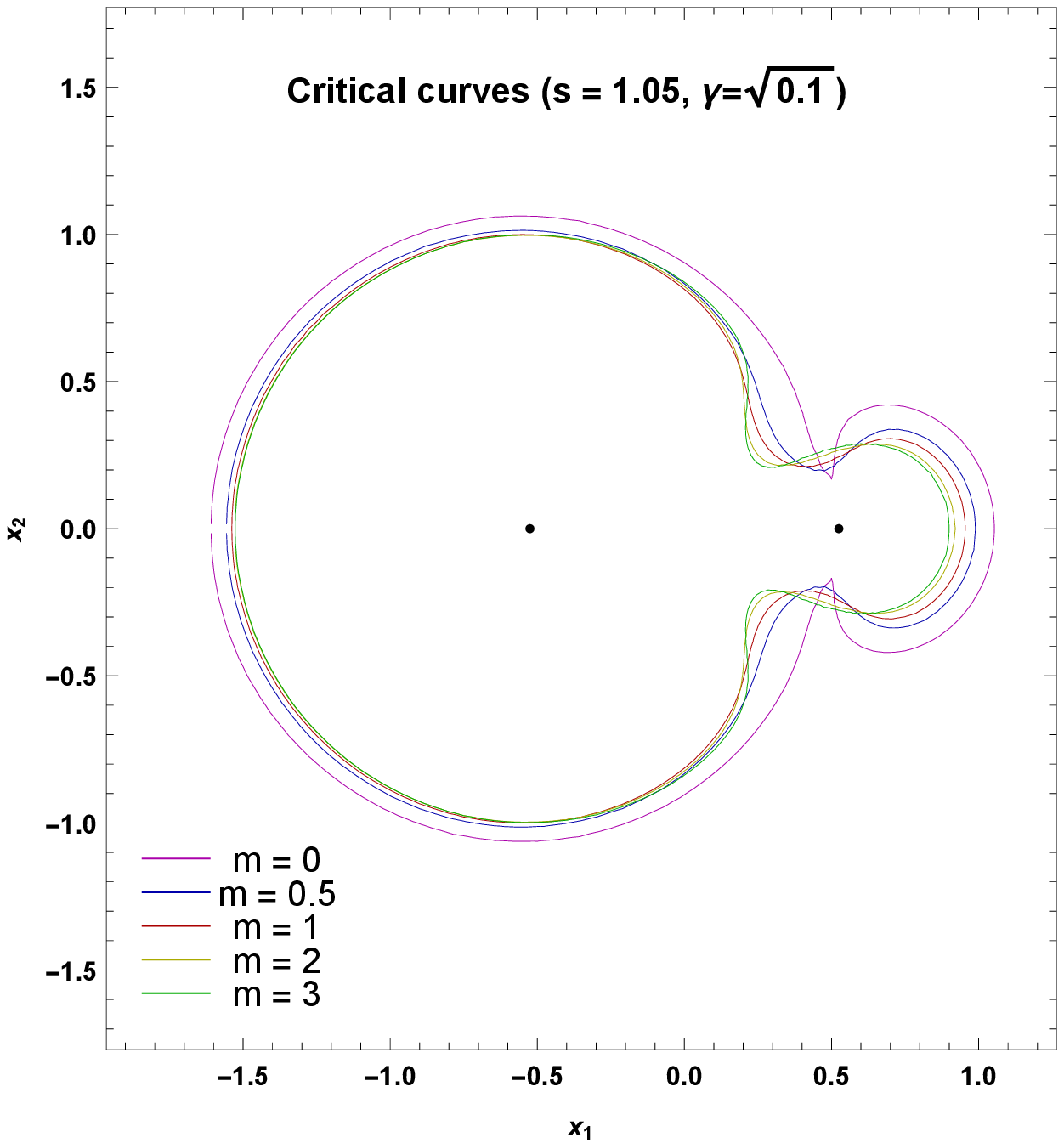}
\includegraphics[height=6.7 cm]{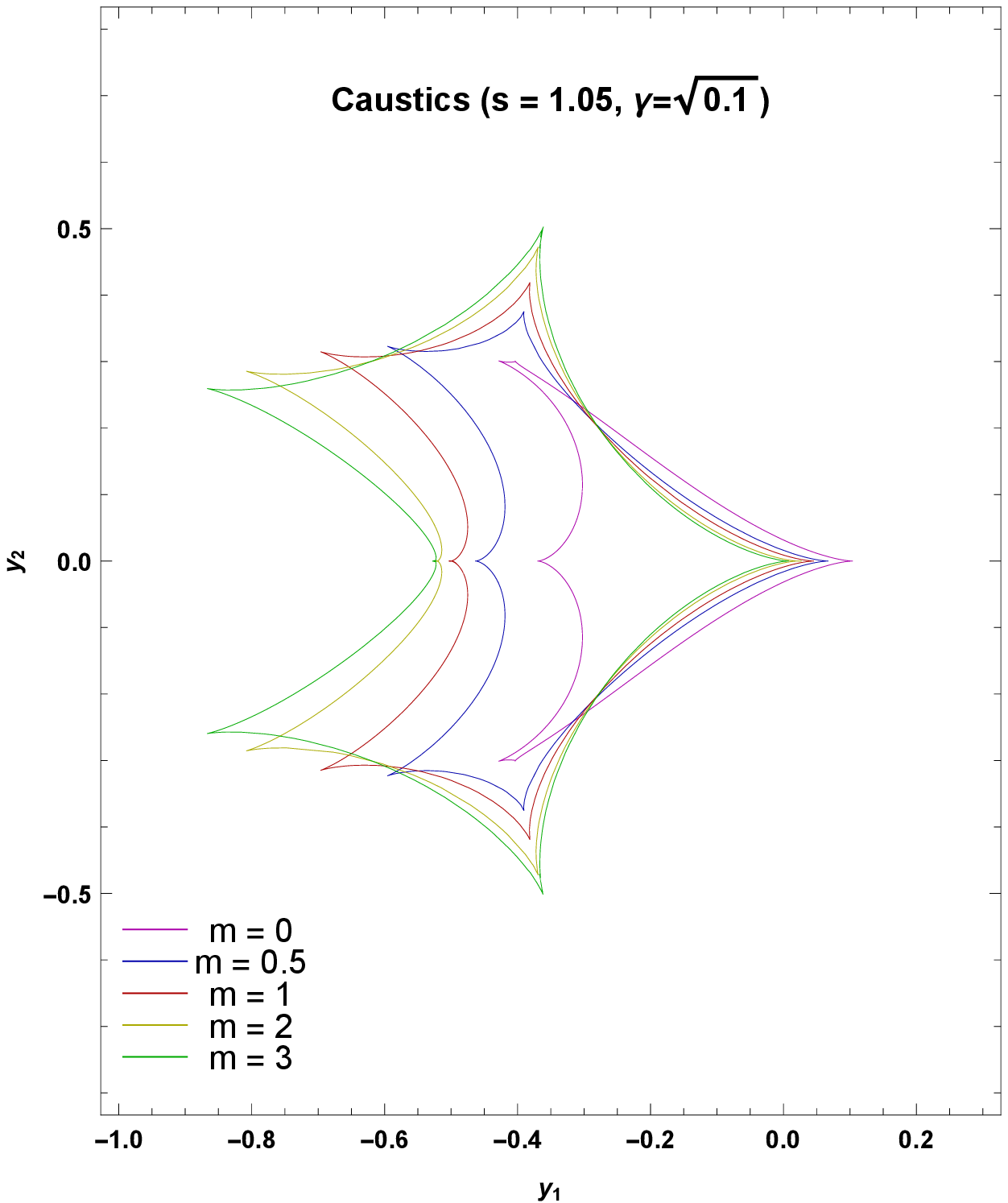}
\caption{Critical curves and caustics in the unequal-strength binary, intermediate separation.}\label{int01}
\end{figure}

In Fig.\ref{int01} we can see the intermediate separation for $s=1.05$. Critical curves now are all joined and they get smaller with increasing $m$.
Caustics now have the 6-cusps shape and they get smaller with decreasing $m$. Note that the throat of the critical curve is wider for $m>2$ and narrower for $m<1$. Correspondingly, the fold between the off-axis cusps is longer for $m>1$ and is extremely short for $m=0$, where the two off-axis cusps almost coincide.

\begin{figure}[H]
\centering
\includegraphics[height=6.7 cm]{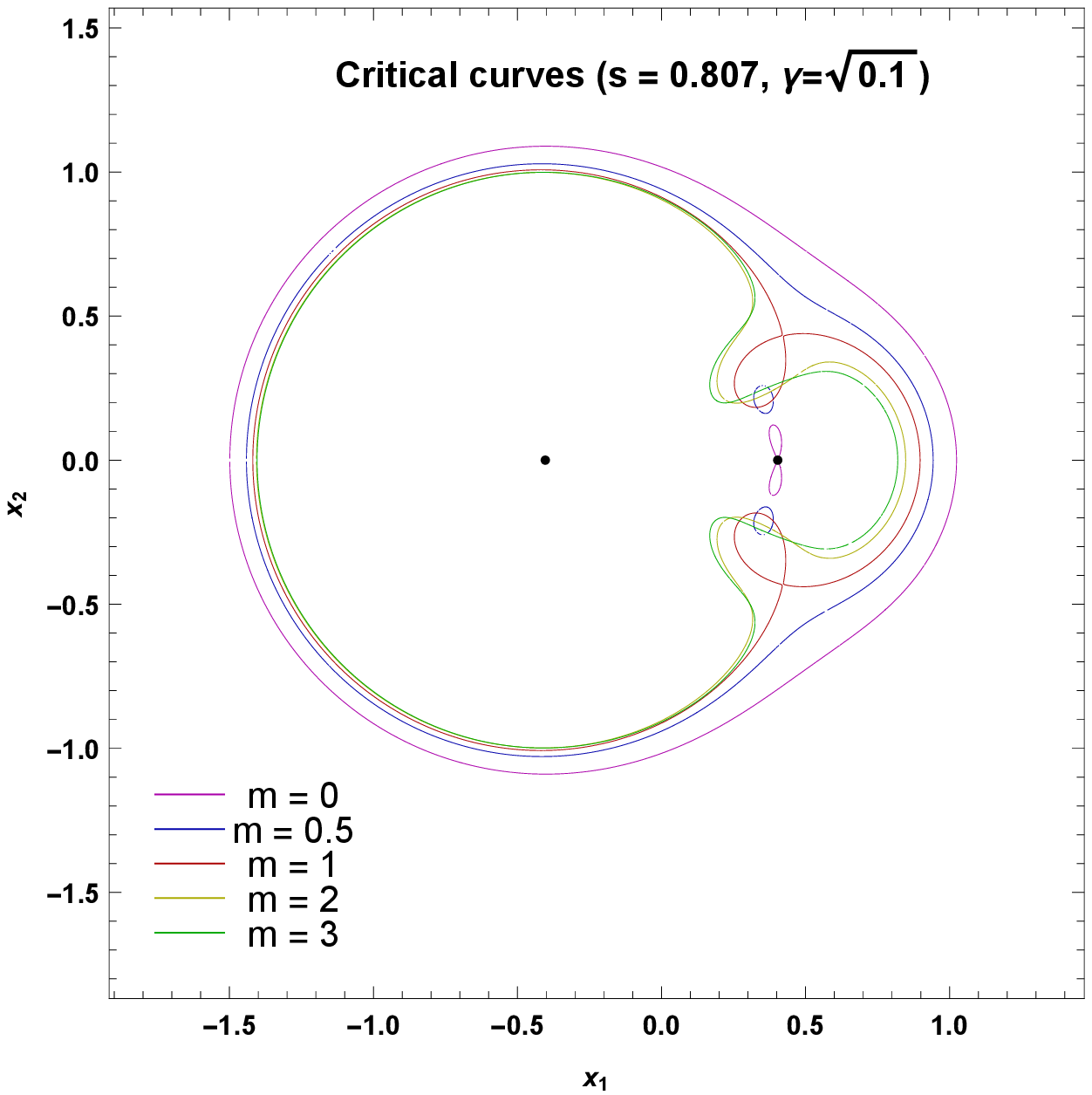}
\includegraphics[height=6.7 cm]{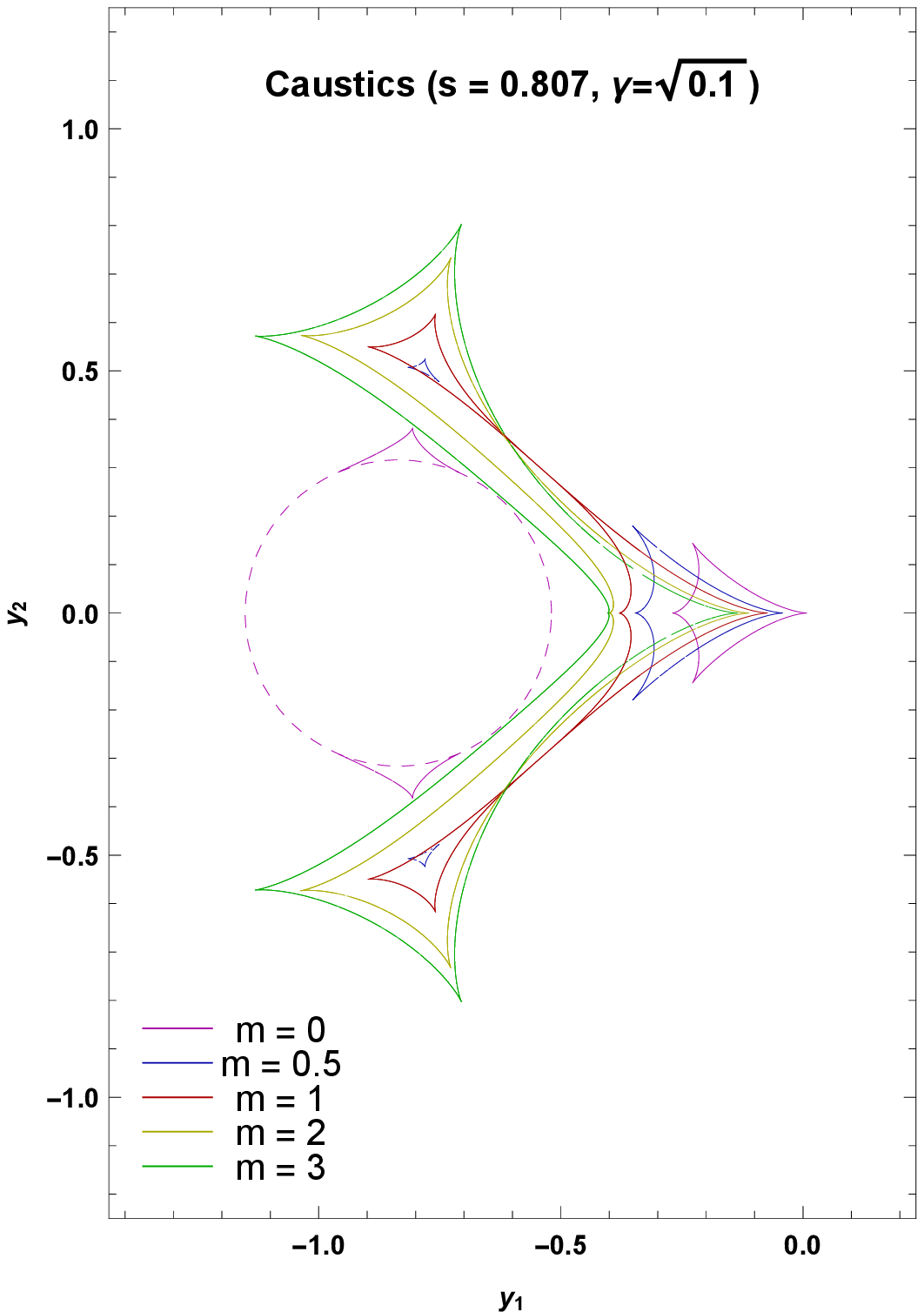}
\caption{Critical curves and caustics in the unequal-strength binary, close-intermediate transition. Dashed magenta circle indicates the pseudocaustic for $m=0$.}\label{01ci}
\end{figure}

In Fig.\ref{01ci} we have the close-intermediate transition for $s=0.807$. For $m=0$ and $m=0.5$, the primary caustics are already in the close regime, with the smaller ovals detached from the primary critical curve; for $m=1$ we see the transition (red line), for $m>1$ we are still in the intermediate regime. Note that the $m=0$ ovals already reached the right side lens and the triangular caustics reached the pseudocaustic.

\begin{figure}[H]
\centering
\includegraphics[height=6.7 cm]{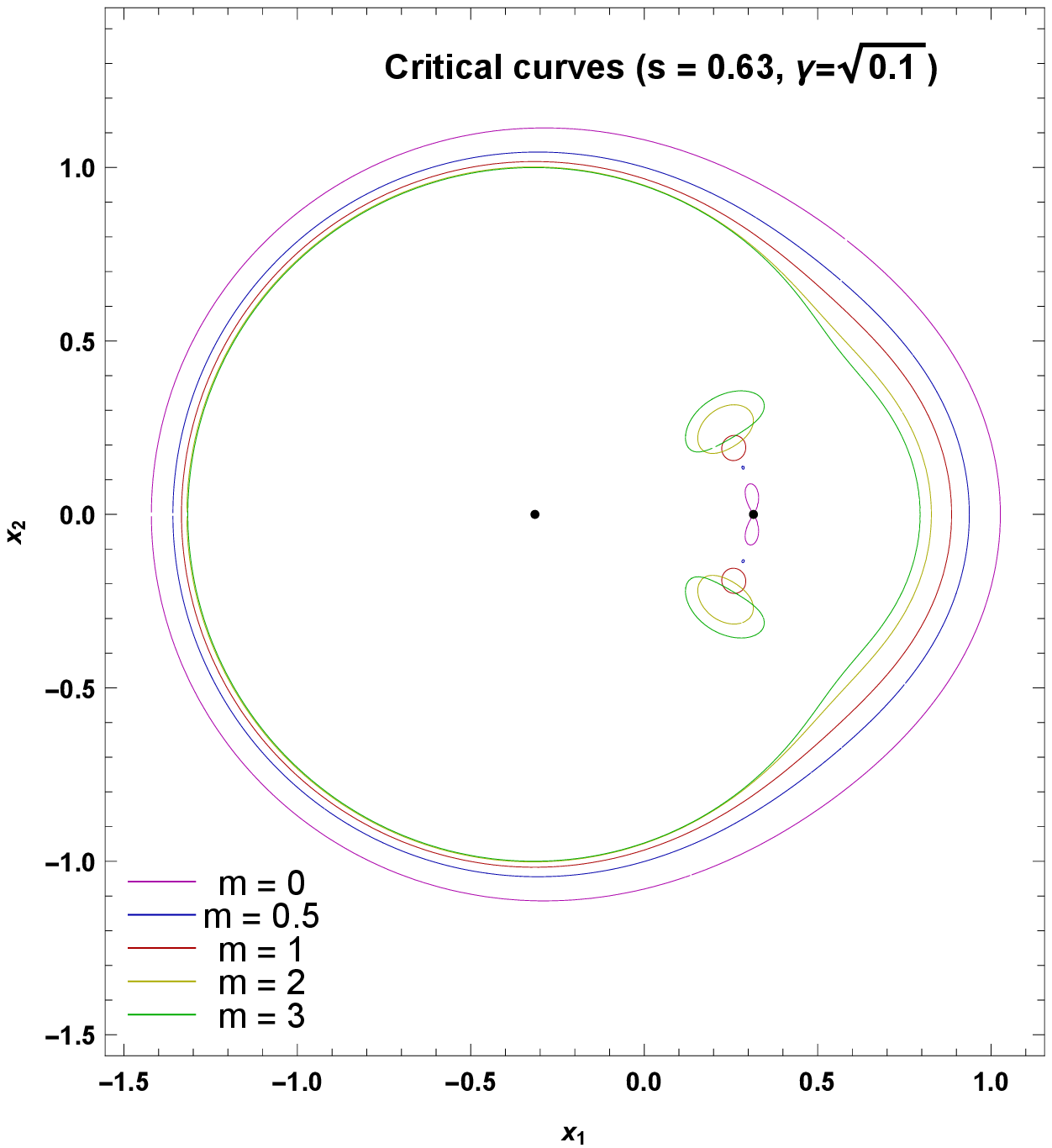}
\includegraphics[height=6.7 cm]{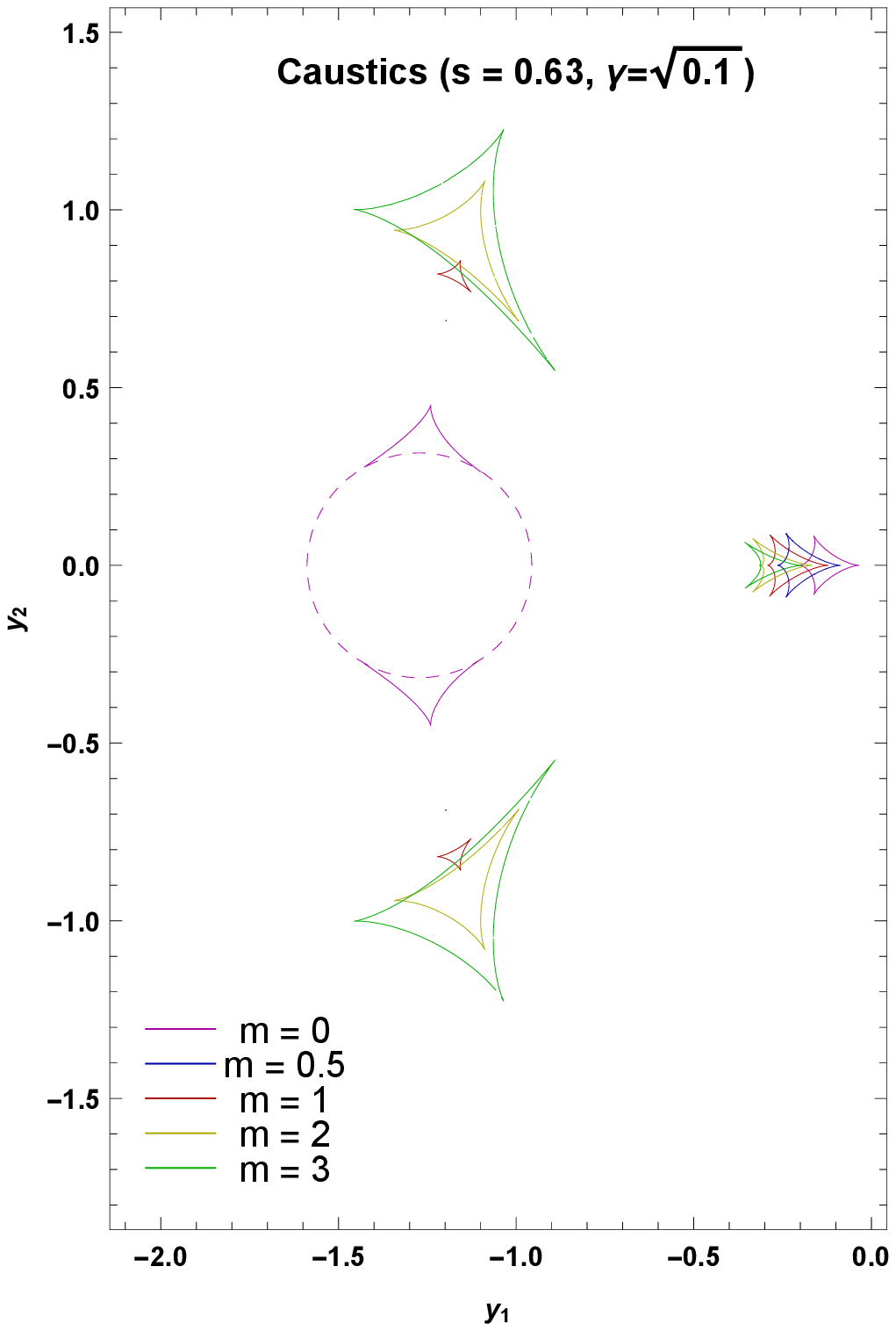}
\caption{Critical curves and caustics in the unequal-strength binary, close separation. The dashed magenta circle indicates the pseudocaustic for $m=0$.}\label{01clo}
\end{figure}

In Fig.\ref{01clo} we show the close separation for $s=0.63$: the main critical curves, that generate the central caustics, are big ovals growing up in size with decreasing $m$. Secondary critical curves are small ovals close to the second lens, moving in the left direction as $m$ increases. Like the equal-strength ratio case, for $m=0$ the secondary critical curves are attached in a lemniscate shape and the corresponding caustics remain open on the pseudocaustic (see subsection \ref{pseudo}).

On the right panel we have the caustics: as $m$ decreases, the central caustic moves to the right; secondary caustics become larger for greater values of $m$.

\subsection{Reversed unequal-strength binary}

In the previous section we assumed that the bigger lens was standard ($n=1$) and the smaller lens had a different index $m$. In this section we study the reverse situation: the standard lens is smaller and the other lens is bigger. We thus keep $\gamma=\sqrt{0.1}$, fix $m=1$ and let $n$ vary.

\begin{figure}[H]
\centering
\includegraphics[height=6.6 cm]{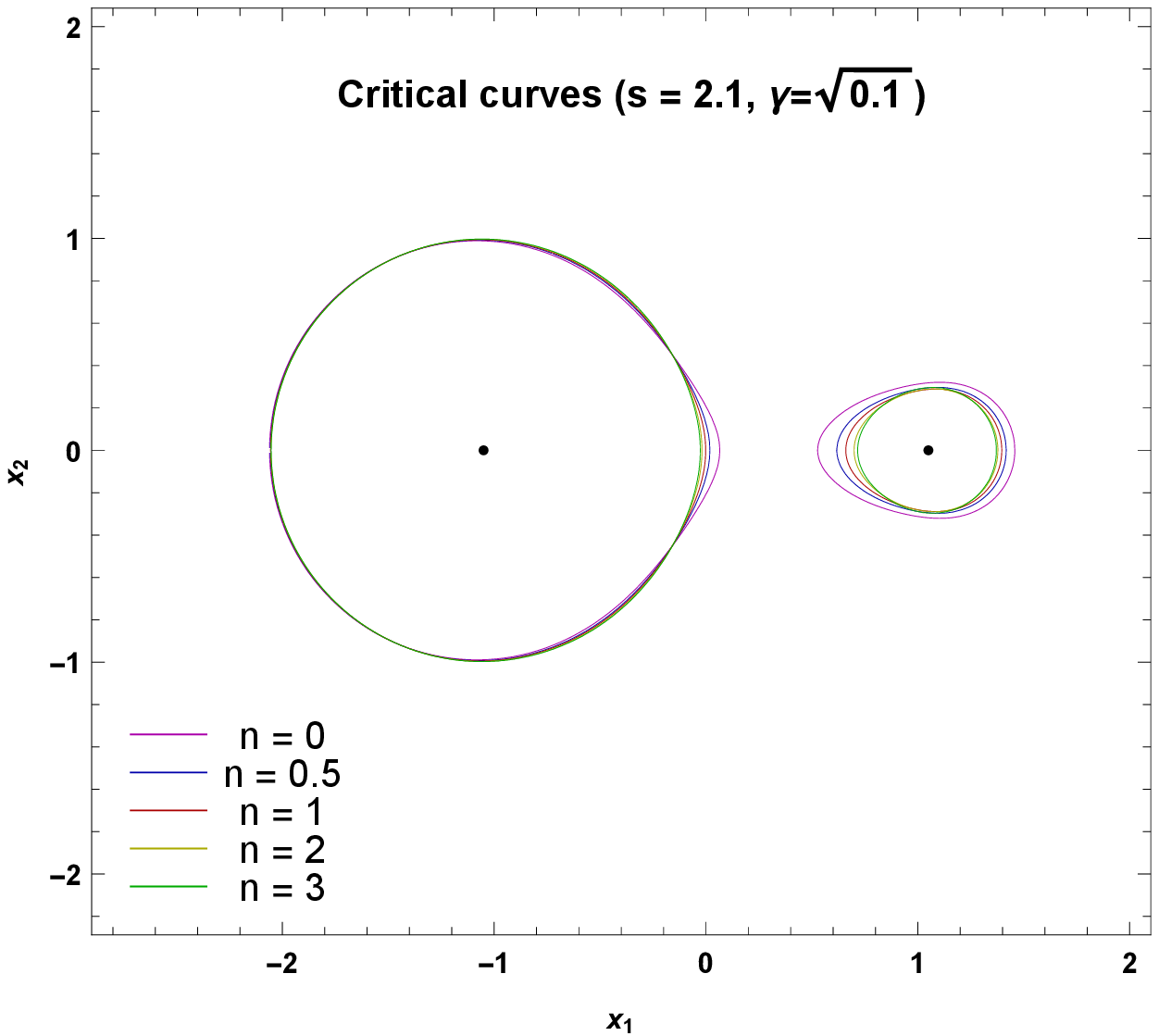}
\includegraphics[height=6.6 cm]{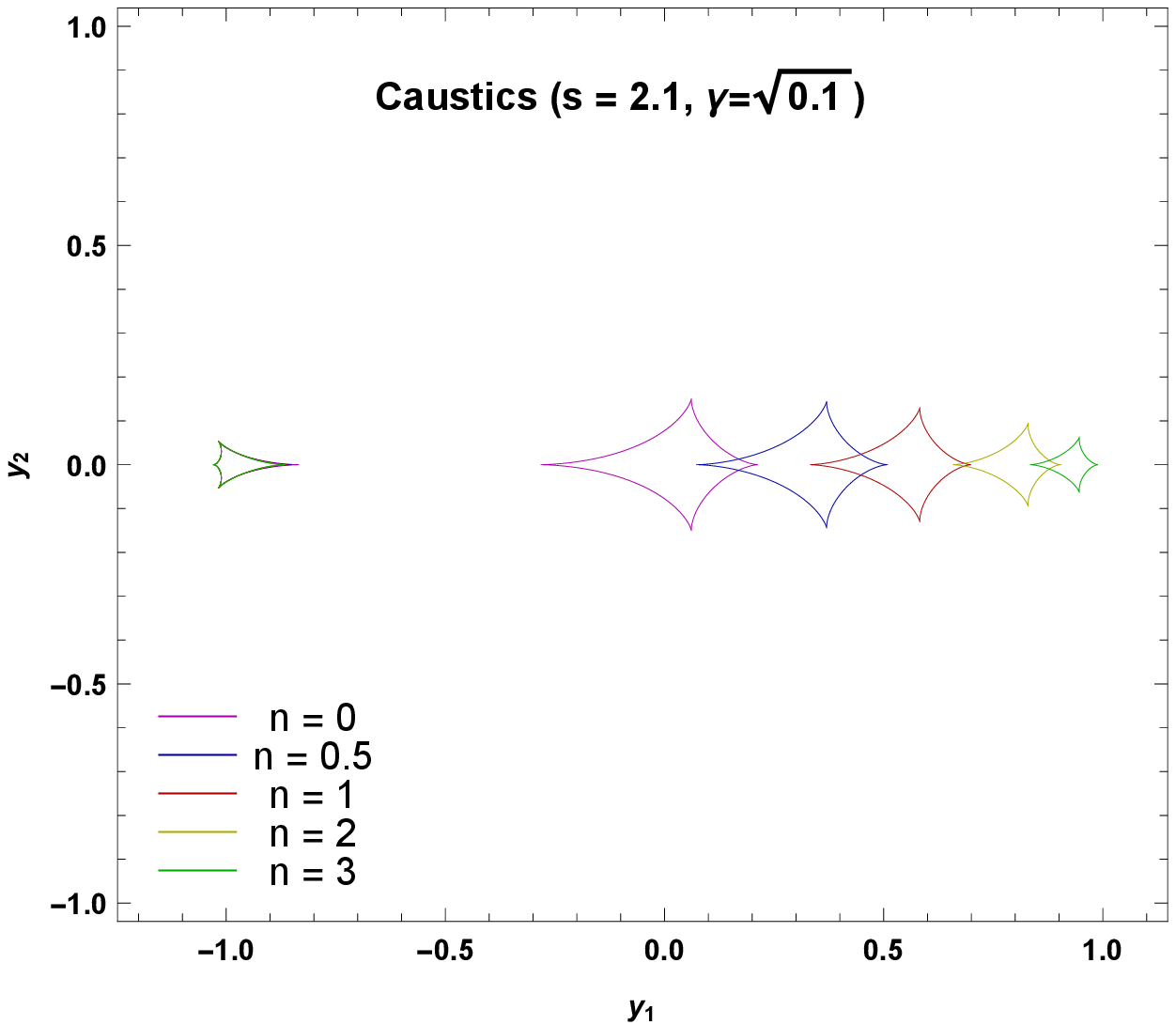}
\caption{Critical curves and caustics in the unequal-strength binary with the standard lens on the right, wide separation. Here and in the following figures the lens on the right has $m = 1$ and the lens on the left has variable $n$.}\label{exwide}
\end{figure}

In Fig.\ref{exwide} we start from the wide separation. Similarly to Fig. \ref{wide01}, the caustic of the non-standard lens remains unaffected, while the caustic of the standard object strongly depends on the tidal field of the other lens. The shift and the size are much more affected than before, since now the standard lens is the weaker one.

\begin{figure}[H]
\centering
\includegraphics[height=6.55 cm]{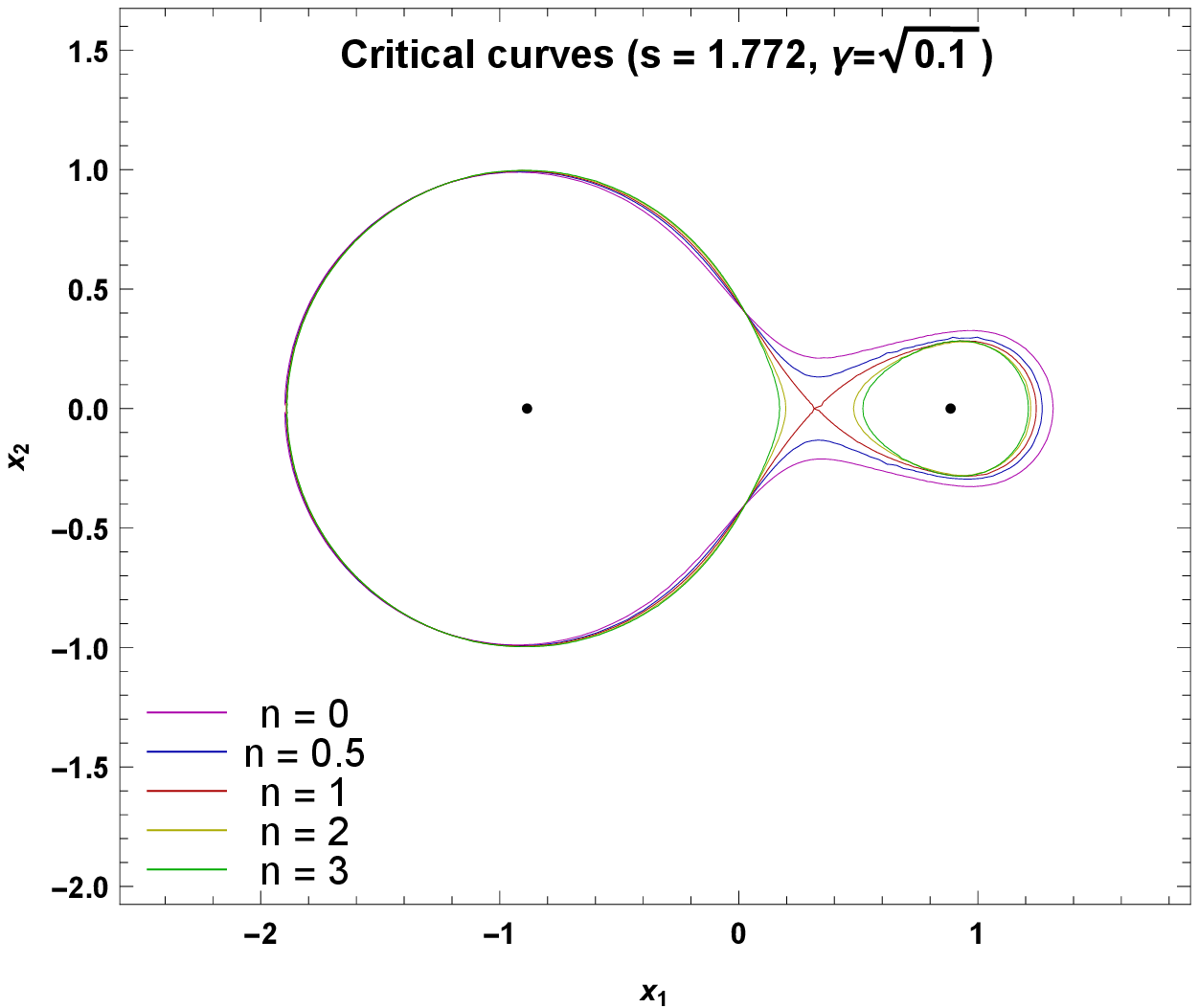}
\includegraphics[height=6.55 cm]{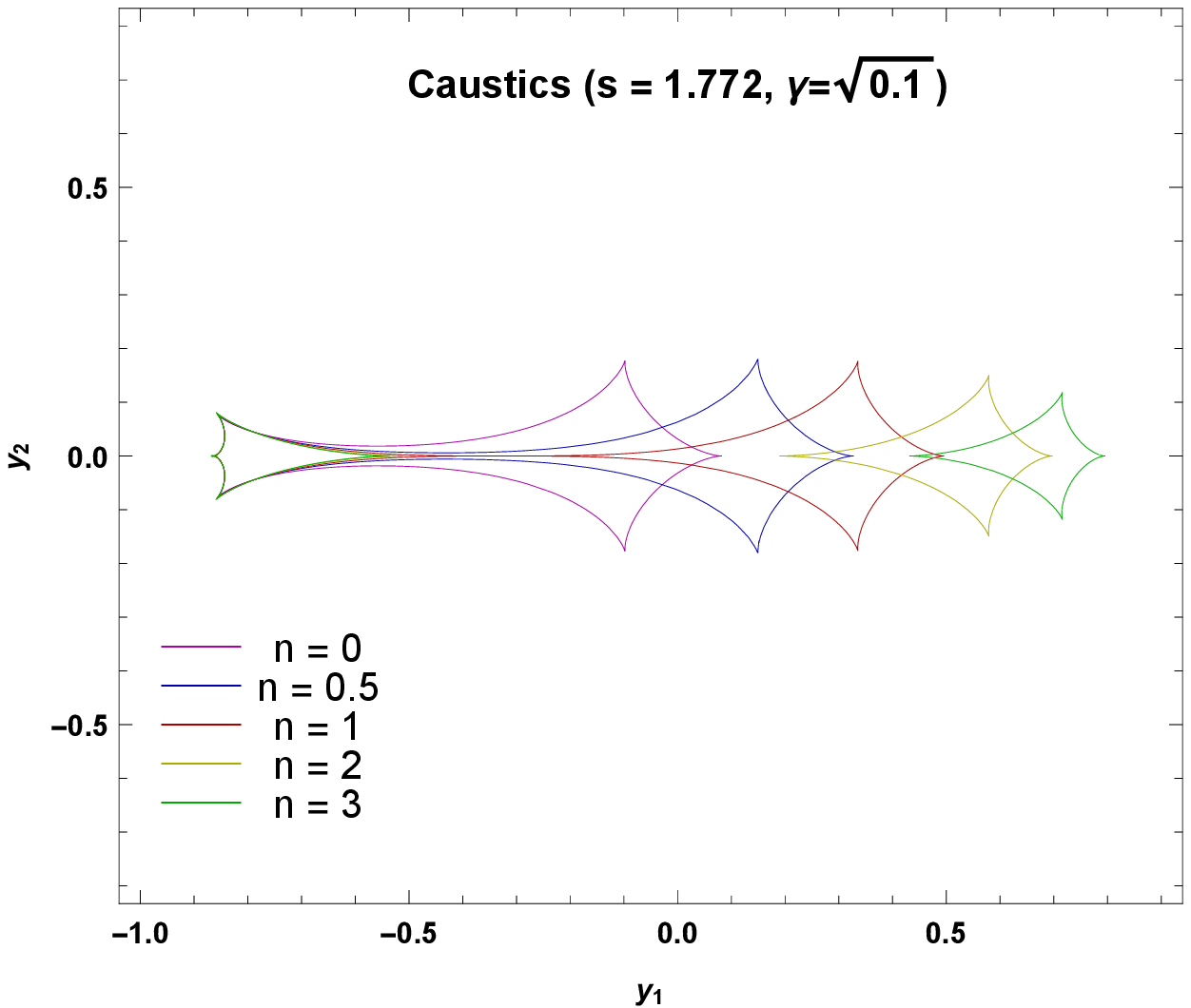}
\caption{Critical curves and caustics in the unequal-strength binary with the standard lens on the right, intermediate-wide transition.}\label{extriw}
\end{figure}

In Fig.\ref{extriw}, we are at the intermediate-wide transition. The situation is quite similar to Fig. \ref{iw01}, with stronger dependence on the index $n$, as discussed before.

\begin{figure}[H]
\centering
\includegraphics[height=6.7 cm]{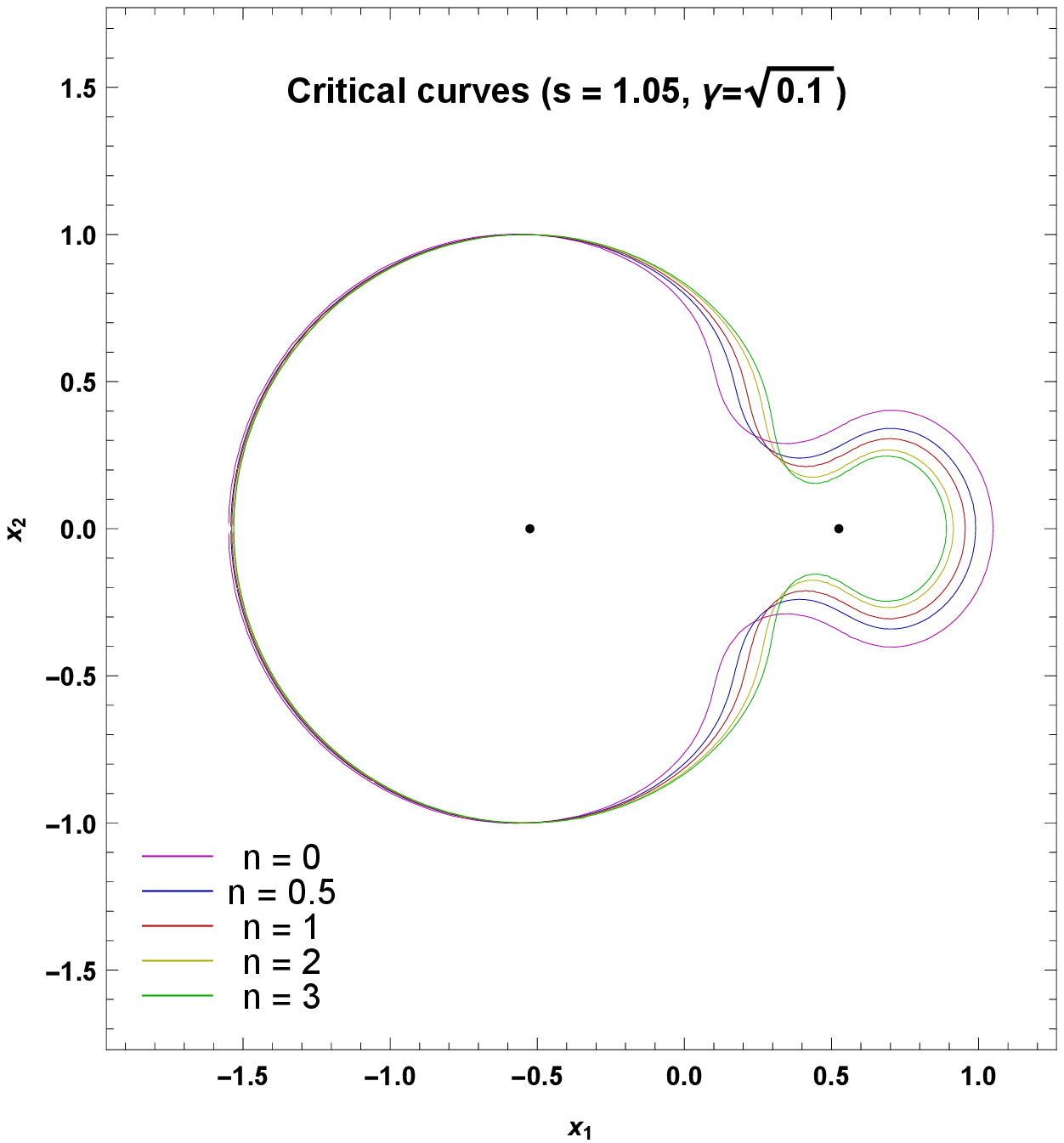}
\includegraphics[height=6.7 cm]{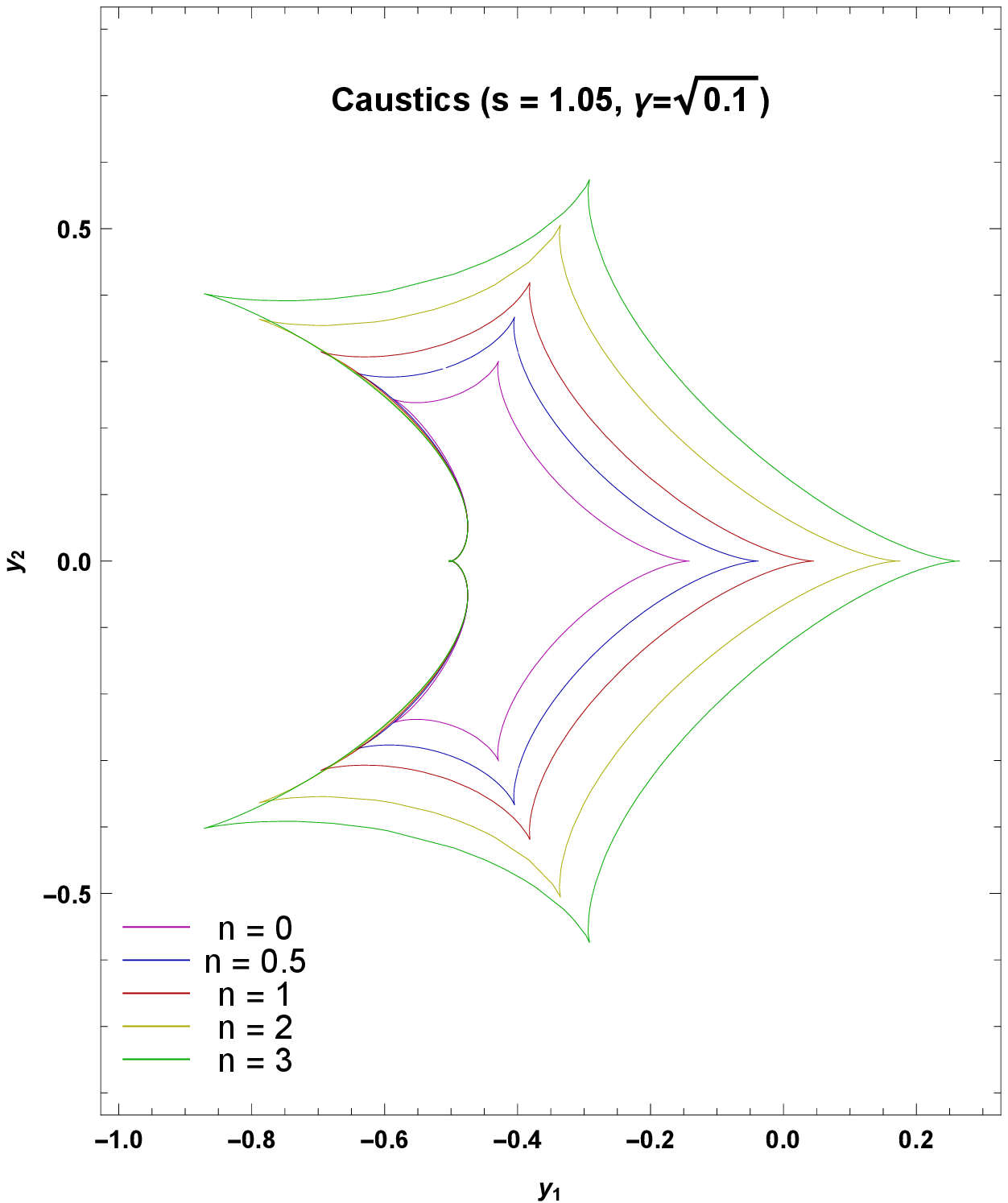}
\caption{Critical curves and caustics in the unequal-strength binary with the standard lens on the right, intermediate separation.}\label{exint}
\end{figure}

Fig. \ref{exint} shows the intermediate topology. Comparing with Fig. \ref{int01}, it is interesting to note that here the left cusp is common for all caustics, while there it was the right cusp to be shared among all caustics. Of course, we can still interpret this fact through the variations of the tidal fields.

\begin{figure}[H]
\centering
\includegraphics[height=6.7 cm]{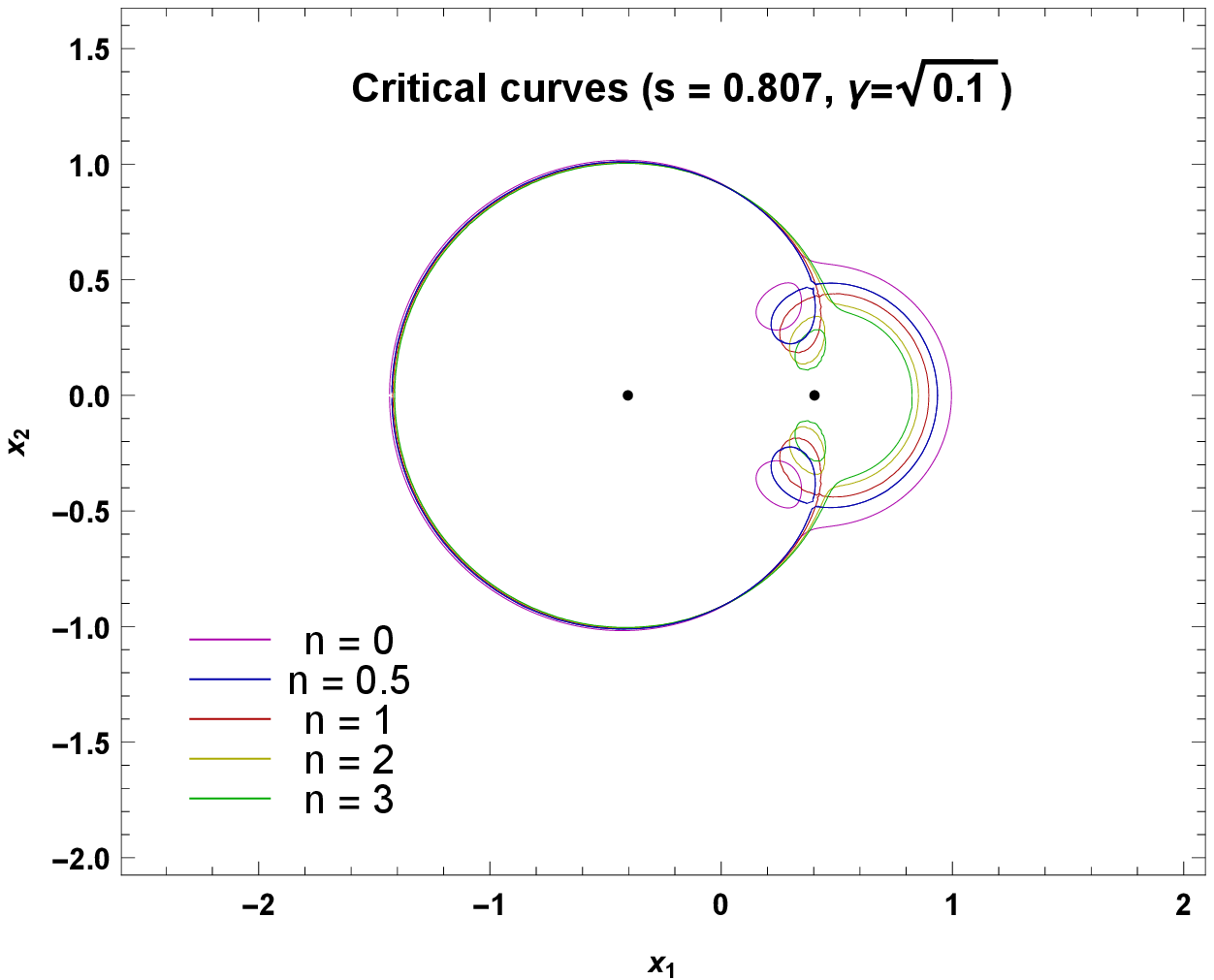}
\includegraphics[height=6.7 cm]{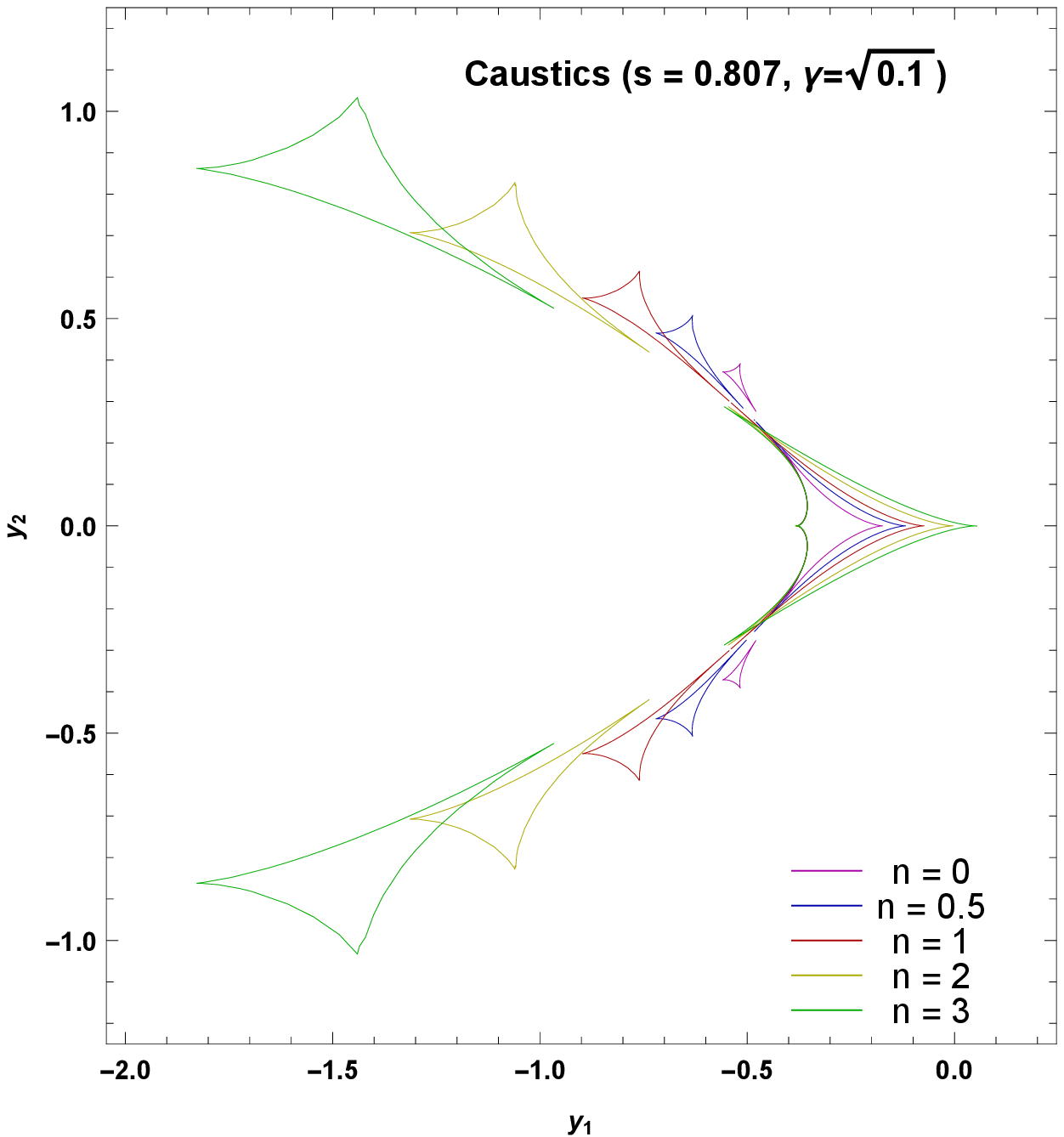}
\caption{Critical curves and caustics in the unequal-strength binary with the standard lens on the right, close-intermediate transition.}\label{extrci01}
\end{figure}

Fig. \ref{extrci01} shows the close-intermediate transition. Note that the red curves ($n=m=1$) are exactly at the transition, while both larger and smaller $n$ curves are in the close regime. This is not what happens in Fig. \ref{01ci}, where larger $m$ curves were still in the intermediate regime. Then we learn that the close regime is more extended for all $n\neq 1$ in this case.

\begin{figure}[H]
\centering
\includegraphics[height=6.7 cm]{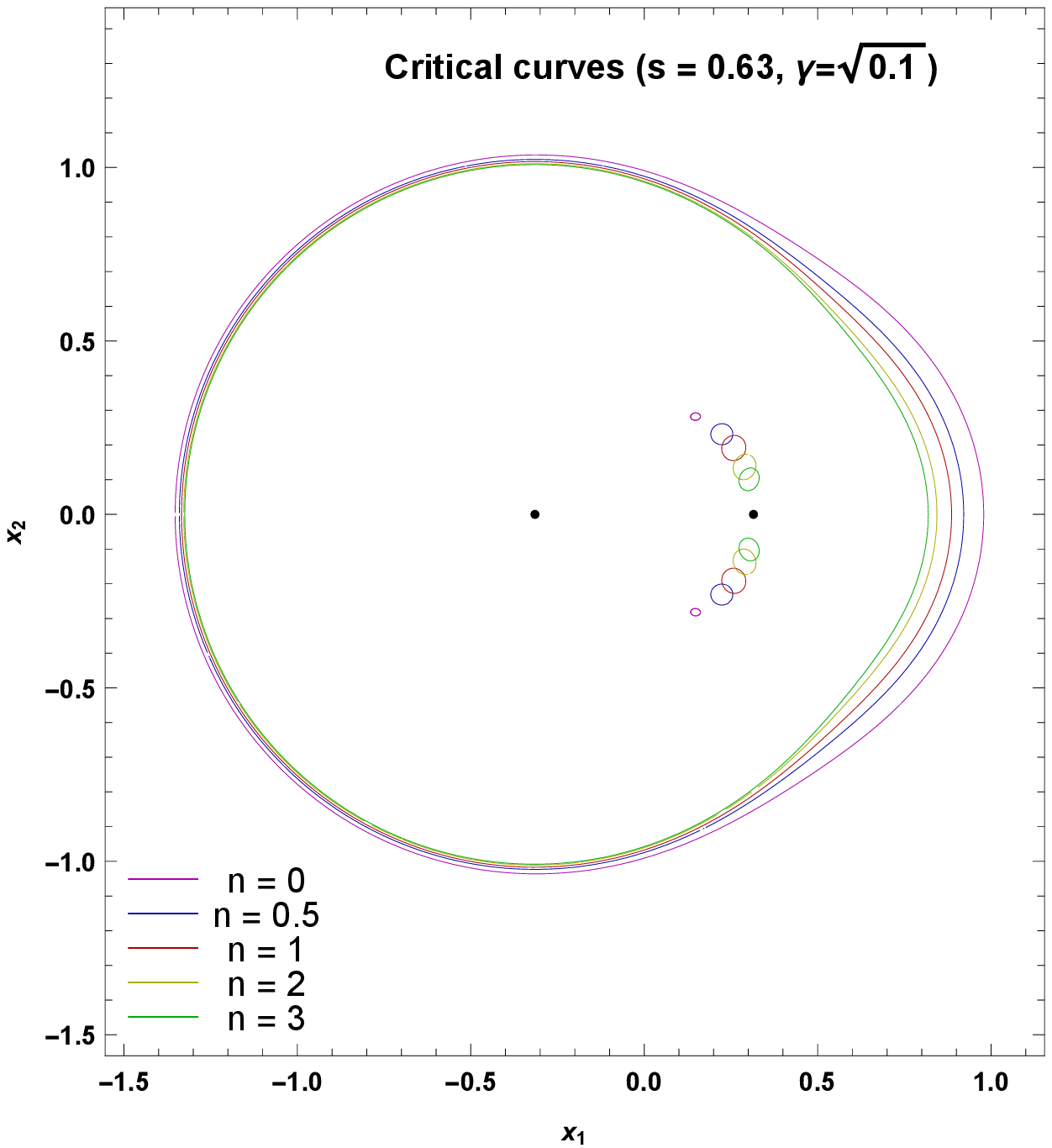}
\includegraphics[height=6.7 cm]{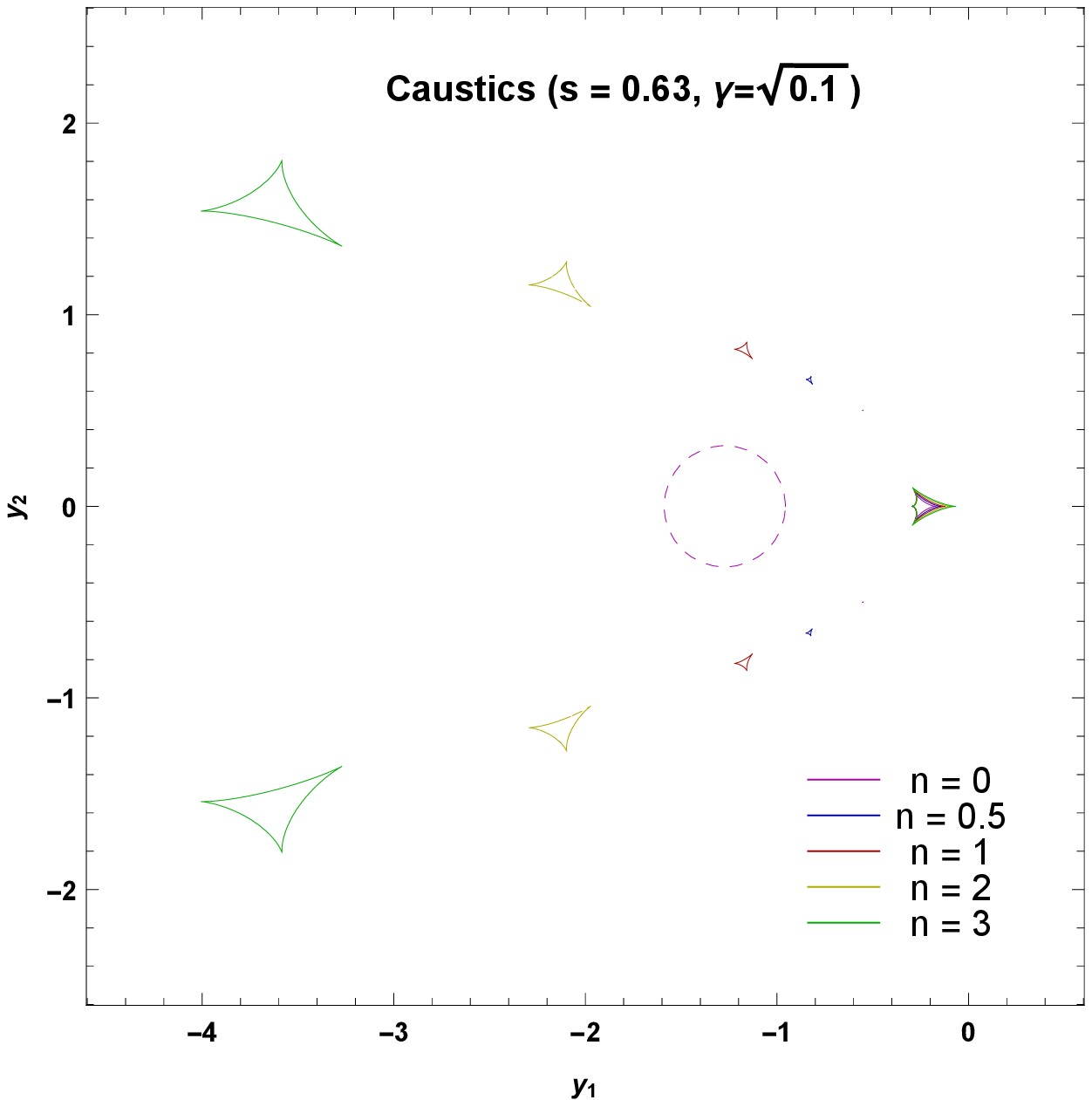}
\caption{Critical curves and caustics in the unequal-strength binary with the standard lens on the right, close separation.}\label{ex01c}
\end{figure}

Finally, Fig. \ref{ex01c} shows the close regime. Note that the $n=0$ small ovals do not collapse to the left lens but remain quite far. The pseudocaustic is never reached by the triangular caustics.

\section{Transitions between different topologies\label{sectra}}

Now we want to find out the boundaries for the three topology regimes, $s_{CI}$ and $s_{IW}$, for any $n$, $m$ and $\gamma$.

As we know that transitions occur via higher order singularities of the lens map (beak-to-beak singularity in the binary lens case), in order to find out $s_{CI}$ and $s_{IW}$ we need to solve again the system of equations 

\begin{equation}
\begin{sistema}
$J=0$ \\
$$\frac{\partial J}{\partial z}=0$$
\end{sistema}
\end{equation}

We put the origin of the system in the first lens, so we rewrite the lens equation as follows

\begin{equation}
 \zeta=z-\frac{1}{\left({z}\right)^{\frac{n-1}{2}} \left(\bar{z}\right)^{\frac{n+1}{2}}}-\frac{\gamma^{m+1}}{\left(z-s\right)^{\frac{m-1}{2}} \left(\bar{z}-s\right)^{\frac{m+1}{2}}}\label{lensq0}
\end{equation}

the Jacobian determinant is 


\begin{equation}
    \begin{split}
J &=\frac{1}{4} \Bigg[\left(2+\frac{\text{n}-1}{z^{\frac{n+1}{2}} \bar{z}^{\frac{n+1}{2}}}+\frac{(m-1) \gamma ^{m+1}}{(z-s)^{\frac{m+1}{2}} (\bar{z}-s)^{\frac{m+1}{2}}}\right)^2-\\
&\left(\frac{\text{n}+1}{z^{\frac{n+3}{2}} \bar{z}^{\frac{n-1}{2}}}+\frac{(m+1) \gamma ^{m+1}}{(z-s)^{\frac{m+3}{2}} (\bar{z}-s)^{\frac{m-1}{2}}}\right) \left(\frac{\text{n}+1}{z^{\frac{n-1}{2}} \bar{z}^{\frac{n+3}{2}}}+\frac{(m+1) \gamma ^{m+1}}{(z-s)^{\frac{m+1}{2}} (\bar{z}-s)^{\frac{m+3}{2}}}\right)\Bigg]\label{J}
    \end{split}
\end{equation}

and


\begin{equation}
\begin{split}
\frac{\partial J}{\partial z} &=        
\frac{1}{4}\Bigg\{\frac{1}{2}\Bigg[ \left(\frac{\text{n}+1}{z^{\frac{n+3}{2}} \bar{z}^{\frac{n+1}{2}}}+\frac{(m+1) \gamma ^{m+1}}{(z-s)^{\frac{m+3}{2}} (\bar{z}-s)^{\frac{m-1}{2}}}\right) \left(\frac{\text{n}^2-1}{z^{\frac{n+1}{2}} \bar{z}^{\frac{n+3}{2}}}+\frac{\left(m^2-1\right) \gamma ^{m+1}}{(z-s)^{\frac{m+1}{2}} (\bar{z}-s)^{\frac{m+3}{2}}}\right)+\\
&\left(\frac{(\text{n}+1) (\text{n}+3)}{z^{\frac{n+5}{2}} \bar{z}^{\frac{n-1}{2}}}+\frac{(m+1) (m+3) \gamma ^{m+1}}{(z-s)^{\frac{m+5}{2}} (\bar{z}-s)^{\frac{m-1}{2}}}\right) \left(\frac{\text{n}+1}{z^{\frac{n-1}{2}} \bar{z}^{\frac{n+3}{2}}}+\frac{(m+1) \gamma ^{m+1}}{(z-s)^{\frac{m-1}{2}} (\bar{z}-s)^{\frac{m+3}{2}}}\right)\Bigg]\\
&-\left(2+\frac{\text{n}-1}{z^{\frac{n+1}{2}} \bar{z}^{\frac{n+1}{2}}}+\frac{(m-1) \gamma ^{m+1}}{(z-s)^{\frac{m+1}{2}} (\bar{z}-s)^{\frac{m+1}{2}}}\right) \left(\frac{\text{n}^2-1}{z^{\frac{n+3}{2}} \bar{z}^{\frac{n+1}{2}}}+\frac{\left(m^2-1\right) \gamma ^{m+1}}{(z-s)^{\frac{m+3}{2}} (\bar{z}-s)^{\frac{m+1}{2}}}\right)\Bigg\} \label{dJ}
\end{split}
\end{equation}

Here we show the analytical procedure to find out $s_{IW}$; the other transition, $s_{CI}$, is only found numerically.

We require $z=\bar{z}$ because the beak-to-beak singularity for the intermediate-wide transition occurs along the line that joins the two lenses, and we introduce two variables 
\begin{equation}
y_1=\frac{(s-z)^{m+1}}{z^{n+1}},y_2=\frac{(s-z)^{m+2}}{z^{n+2}}\label{y1y2}
\end{equation}

we replace $y_1$ in Eq. (\ref{J}), we solve and we get

\begin{equation}
 y_1=\frac{z^{n+1}-1}{\gamma ^{m+1}},\label{y1y2b}
\end{equation}

we substitute $y_2$ in Eq. (\ref{dJ}), we solve and we find

\begin{equation}
y_2 = \frac{n+1}{(m+1) \gamma ^{m+1}}\label{y1y2bis}.   
\end{equation}

We use Eqs. (\ref{y1y2b}) and (\ref{y1y2bis}) in Eqs. (\ref{y1y2}), we find two new equations and by a combination of them we get a complicated expression for $z$

\begin{equation}
   \frac{\gamma ^{\frac{m+1}{n+2}} \left(1+\gamma ^{\frac{n+1}{n+2}}\right)^{m+1}\bigg[z+\left(\frac{m+1}{n+1}\right)^{\frac{1}{m+2}}\gamma^{\frac{m+1}{m+2}} z^{\frac{n+2}{m+2}}\bigg]^{n-m}}{\bigg[z+\left(\frac{m+1}{n+1}\right)^{\frac{1}{m+2}}\gamma ^{\frac{m+1}{m+2}}  z^{\frac{n+2}{m+2}}\bigg]^{n+1}-\left(1+\gamma ^{\frac{n+1}{n+2}}\right)^{n+1}}=1\label{z}
\end{equation}
that we can solve only numerically. We call this numerical solution $z_{IW}$.  Finally we get the value of the intermediate-wide transition for general $n$ and $m$
\begin{equation}
    s_{IW}= z_{IW}+ \left(\gamma ^{m+1} z_{IW}^{n+2}\frac{m+1}{n+1}\right)^\frac{1}{m+2}.\label{siw}
\end{equation}

\begin{figure}[H]
\centering
\includegraphics[height=6.85 cm]{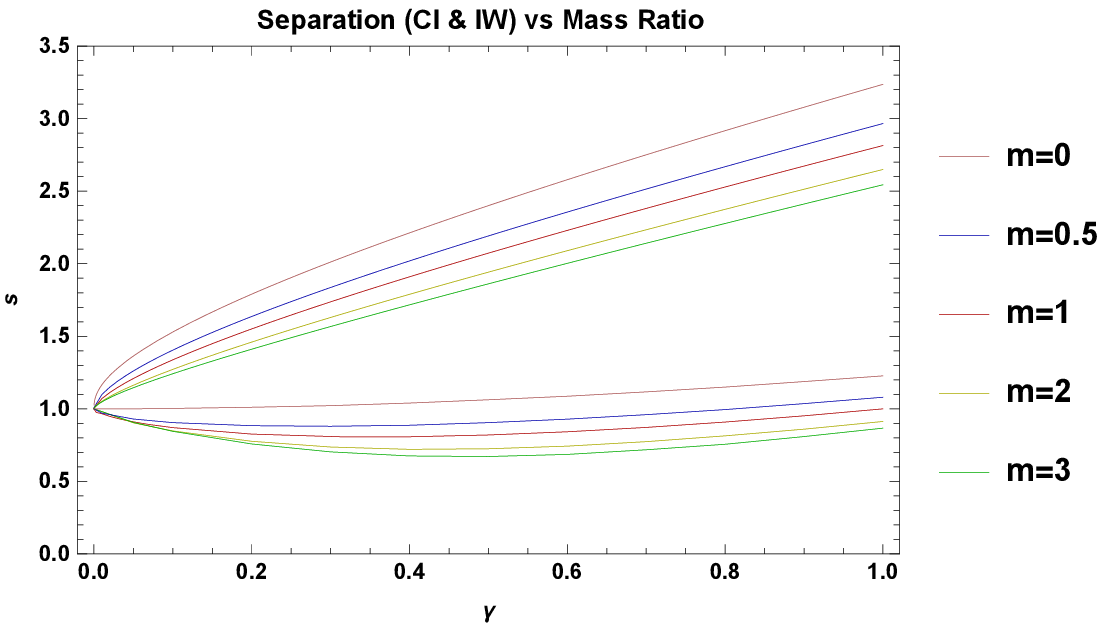}
\quad
\includegraphics[height=6.85 cm]{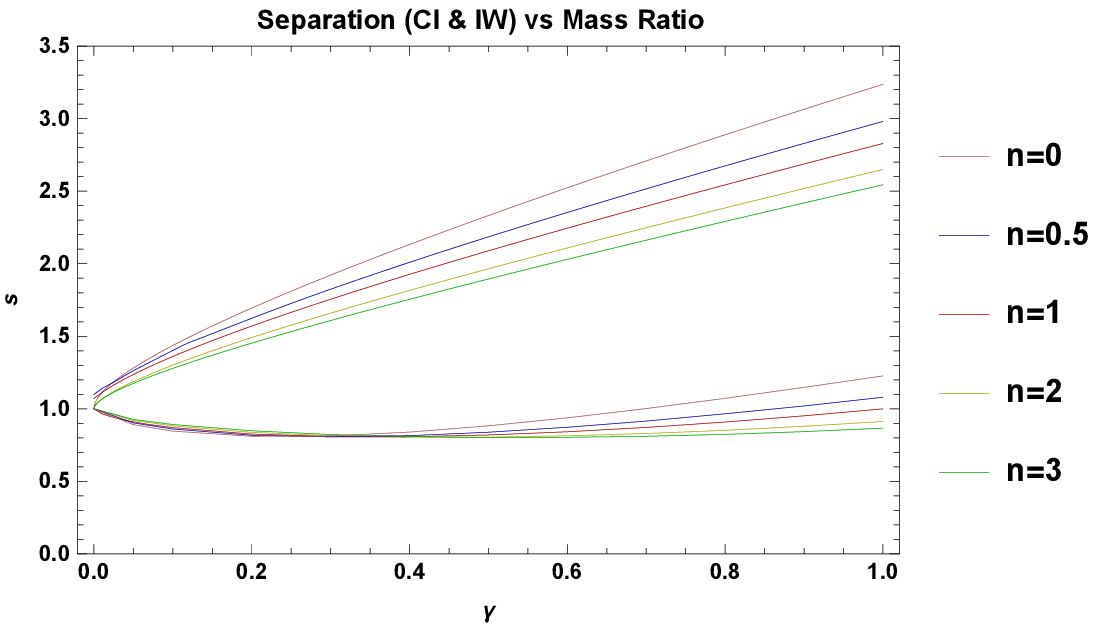}
\caption{Critical values of the separation for the intermediate-wide transition $s_{IW}$ as a function of $\gamma$ (upper curves); critical values of the separation for the close-intermediate transition $s_{CI}$ as a function of $\gamma$ (lower curves). The upper panel is for $n=1$ and variable $m$; the lower panel is for $m=1$ and variable $n$. }\label{transitions}
\end{figure}

In Fig. \ref{transitions}, upper panel, we plot the cases with fixed $n=1$ and variable $m$ (upper curves). The close-intermediate transition $s_{CI}$ is found numerically (lower curves). We can see that the value of $s_{IW}$ increases with $\gamma$ and that the transition occurs earlier for greater values of $m$. The value of $s_{CI}$ has a different behaviour: first it decreases with increasing $\gamma$, with a minimum around $\gamma=0.5$, and then it starts to grow up again. Also in this case the transition occurs earlier for greater values of $m$. We remind the reader that we are working in units of the Einstein radius of the first lens.

For the reversed binary case, in Fig. \ref{transitions}, lower panel, we plot the cases with fixed $m=1$ and vary $n$. We can see that the value of $s_{IW}$ increases with $\gamma$ similarly to the case with fixed $m$. For $s_{CI}$ all curves are very closely packed and have a minimum for a value of $\gamma$ that depends on the specific choice of $n$. In particular, for $\gamma=\sqrt{0.1}$, corresponding to the situation in Fig. \ref{extrci01}, the transition occurs for $n=1$ at smaller separation than for all other curves.

\section{Analytical Approximations}

In order to remark the differences with the Schwarzschild case $n=m=1$ and to have a deeper understanding in the caustics evolution, we now want to explore the analytical approximations obtained for the most general case, varying both $n$ and $m$. In particular, we will investigate the wide binary regime and the very small $\gamma$ regime, but we are not able to get analytical results for the close binary regime. The main difficulty comes from the fact that the starting point of the expansion would have the two lenses coinciding in the origin, but the resulting Einstein radius can only be calculated numerically. Therefore, even the zero order is not analytic.

\subsection{Wide Binary}

Let us consider the wide binary regime with $s>>1$: the case in which an isolated object is perturbed by another one at a distance much greater than the Einstein radius $\theta_E$.

We can set the origin of our system in the first lens ($z_A=0$ and $z_B=s$), so we can use the same lens equation written in Eq. (\ref{lensq0}) and the Jacobian determinant in Eq. (\ref{J}).

The perturbing object deforms the circular critical curve of the main object with radius $\rho$ and we can find out this deformation $\delta$ through a perturbative approach.

We set

\begin{equation}
    z= \rho(1+\delta ) e^{i\theta}\label{zro}
\end{equation}
and we substitute in $J$.
In our case $\rho=1$ (because for $s\to\infty$ the radius of the critical curves is the Einsten radius, which, in our case, is $\theta_{E,A}=1$), and take $\delta = O(1/s^{m+1})$. We perform a power series expansion for $J$ about the zero point with respect to $1/s$ at first order. Then, we solve $J=0$ and find the correction to the critical curve
\begin{equation}
  \delta= \frac{\gamma ^{m+1} [1-m+(m+1)\cos (2 \theta )]}{2 (n+1) s^{m+1}}
\end{equation}

Now we put Eq. (\ref{zro}) in Eq. (\ref{lensq0}) and we expand around zero with respect to $1/s$ at first order, and we take the real and the imaginary parts

\begin{equation}
    Re[\zeta  (\theta )]= \frac{\gamma ^{m+1} }{s^m}+\frac{(m+1)\gamma ^{m+1}\cos ^3(\theta )}{s^{m+1}} \label{widecau1}
\end{equation}

\begin{equation}
    Im[\zeta  (\theta )]=-\frac{(m+1)  \gamma ^{m+1}  \sin ^3(\theta )}{s^{m+1}} \label{widecau2}
\end{equation}

The real part contains the shift $\frac{\gamma ^{m+1}}{s^m}$ of the caustic toward the direction of perturbing object as we can see in Figs. (\ref{wi}), (\ref{wide01}), (\ref{exwide}).

The other term, $\cos ^3(\theta )+i\sin ^3(\theta )$, describes the shape of the caustic (the 4-cusps astroid) that remains unchanged by varying $m$. The coefficient $\gamma^{m+1}\frac{(m+1)}{s^{m+1}}$ gives the size of the caustic. 

In these approximations the next-to-leading order is given by a term $O(1/s^{2(m+1)})$. So, for any values of $m$ the relative importance of these neglected corrections to the analytic caustics of Eqs. (\ref{widecau1}-\ref{widecau2}) is $1/s^{m+1}$, which means that for $s=10$ and $m=1$ we get a $1\%$ error, while for $m=0$ we get a $10\%$ error. In fact, caustics with $m=0$ are more heavily affected by tidal fields.

\subsection{Extremely unequal-strength ratio limit}

Now we study the caustic evolution in the extreme limit $\theta_{E,B}<<\theta_{E,A}$ for the close and wide separations. We remind that, in the case of two Schwarzschild objects ($n=m=1$) this is the so-called ``planetary'' limit.

\subsubsection{Central Caustic}

We put the origin of our system in the first lens ($z_A=0$), the perturbing object is at $z_B=-s$ and we rewrite the lens equation as follows

\begin{equation}
   \zeta=z-\frac{1}{\left({z}\right)^{\frac{n-1}{2}} \left(\bar{z}\right)^{\frac{n+1}{2}}}-\frac{\gamma^{m+1}}{\left(z+s\right)^{\frac{m-1}{2}} \left(\bar{z}+s\right)^{\frac{m+1}{2}}}.\label{lensqGamma} 
\end{equation}

The Jacobian determinant is

\begin{equation}
    \begin{split}
J &=\frac{1}{4} \Bigg[\left(2+\frac{\text{n}-1}{z^{\frac{n+1}{2}} \bar{z}^{\frac{n+1}{2}}}+\frac{(m-1) \gamma^{m+1}}{(z+s)^{\frac{m+1}{2}} (\bar{z}+s)^{\frac{m+1}{2}}}\right)^2-\\
&\left(\frac{\text{n}+1}{z^{\frac{n+3}{2}} \bar{z}^{\frac{n-1}{2}}}+\frac{(m+1) \gamma^{m+1}}{(z+s)^{\frac{m+3}{2}} (\bar{z}+s)^{\frac{m-1}{2}}}\right) \left(\frac{\text{n}+1}{z^{\frac{n-1}{2}} \bar{z}^{\frac{n+3}{2}}}+\frac{(m+1) \gamma^{m+1}}{(z+s)^{\frac{m+1}{2}} (\bar{z}+s)^{\frac{m+3}{2}}}\right)\Bigg]\label{JG}
    \end{split}
\end{equation}

For the critical curve of the main lens, we use the parametrization in $J$

\begin{equation}
    z= (1+  \delta )e^{i \theta }
\end{equation}
where $\delta=O(\gamma^{m+1})$.

We expand around zero with respect to $\gamma^{m+1}$ to first order, solve $J=0$ and we find the correction of the circular critical curve

\begin{equation}
   \delta= \frac{2+4 s \cos \theta +s^2 [(m+1) \cos (2 \theta )-m+1]}{2 (n+1) \left(1+2 s \cos \theta +s^2\right)^{\frac{m+3}{2}}}\gamma^{m+1}.
\end{equation}

We can substitute this $\delta$ in Eq. \ref{lensqGamma} and obtain the caustic. Since it is not a simple expression, we omit it here.

In order to find the size of the caustic, we evaluate it for $\theta=0$ and $\theta=\pi$. We have

\begin{equation}
    \Delta \zeta = \zeta(0)-\zeta(\pi)= s \gamma^{m+1}   \bigg[\frac{1}{(s-1)^{m+1}}-\frac{1}{(s+1)^{m+1}}\bigg]\label{size}
\end{equation}
and this is the distance between the left and the right cusp. 

We also find that the caustic is invariant under the transformation

\begin{equation}
    s\to\frac{1}{s}, \gamma^{m+1}\to\frac{\gamma^{m+1}}{s^{m-1}}.
\end{equation}
which expresses the duality of the close-wide regimes in our mixed binary framework.

\subsubsection{Caustics of the Perturbing Object}

We put the origin of our system in the second lens so that $z_A=-s$ and $z_B=0$ and the lens equation becomes

\begin{equation}
   \zeta=z-\frac{1}{\left({z+s}\right)^{\frac{n-1}{2}} \left(\bar{z}+s\right)^{\frac{n+1}{2}}}-\frac{\gamma^{m+1}}{z^{\frac{m-1}{2}} \bar{z}^{\frac{m+1}{2}}}\label{lensqGamma2}   
\end{equation}

we rewrite the Jacobian determinant as follows

\begin{equation}
    \begin{split}
J &=\frac{1}{4} \Bigg[\left(2+\frac{\text{n}-1}{(z+s)^{\frac{n+1}{2}} (\bar{z}+s)^{\frac{n+1}{2}}}+\frac{(m-1) \gamma^{m+1}}{(z)^{\frac{m+1}{2}} (\bar{z})^{\frac{m+1}{2}}}\right)^2-\\
&\left(\frac{\text{n}+1}{(z+s)^{\frac{n+3}{2}} (\bar{z}+s)^{\frac{n-1}{2}}}+\frac{(m+1) \gamma^{m+1}}{z^{\frac{m+3}{2}} \bar{z}^{\frac{m-1}{2}}}\right) \left(\frac{\text{n}+1}{(z+s)^{\frac{n-1}{2}} (\bar{z}+s)^{\frac{n+3}{2}}}+\frac{(m+1) \gamma^{m+1}}{z^{\frac{m+1}{2}} \bar{z}^{\frac{m+3}{2}}}\right)\Bigg]\label{J1}
    \end{split}
\end{equation}
and we introduce a new expression for $z$

\begin{equation}
z=\rho^{\frac{1}{m+1}}\gamma e^{i\theta}.
\end{equation}

We substitute in Eq. (\ref{J1}) and we expand with respect to $\gamma^{m+1}$, around zero at zero order and the Jacobian determinant becomes
\begin{equation}
  \frac{(\rho -1) (\rho+m )}{\rho ^2}+  \frac{(n-1) (m+2 \rho -1)-(m+1) (n+1) \cos (2 \theta )}{2 \rho  s^{2}}-\frac{n  }{s^{2 n+2}}=0\label{J=0}
\end{equation}
Then we solve Eq. (\ref{J=0}), $J=0$, with respect to $\rho$ and we find two solutions
\begin{equation}
\rho_{\pm}= \frac{(m-1) \{ s^{n+1}\big[ (1-n)+(m+1) (n+1) \cos (2 \theta )-2s^{n+1}\big]\pm\sqrt{\Delta }\}}{4 \left(s^{n+1}-1\right) \left(s^{n+1}+n\right)}
\end{equation}
\begin{equation}
\Delta =s^{2 n+2}\{\big[(m-1) (2 s^{n+1}+n-1)-(m+1) (n+1) \cos (2 \theta )\big]^2+16 m (s^{n+1}-1) (s^{n+1}+n)\}
\end{equation}

We have two scenarios: for external objects (when the secondary lens is outside the Einstein ring of the main lens, $s>1$) the critical curves are elongated rings, see Fig. (\ref{exwide}); for internal objects (when the secondary lens is inside the critical curve of the main lens, $s<1$) it generates two specular ovals, see Fig. \ref{ex01c}. 

In order to get the caustics we put our solutions in the lens equation and we get, at first order:

\begin{equation}
 \zeta=\gamma \rho ^{\frac{1}{m+1}} \bigg[e^{i \theta } \left(\frac{n-1}{2 s^{n+1}}-\frac{1}{\rho }+1\right)+\frac{e^{-i \theta } (n+1)}{2 s^{n+1}}\bigg]-\frac{1}{s^{n}}\label{lensGammazro}     
\end{equation}

From Eq. (\ref{lensGammazro}) we can get all the information for the size and for the displacement of the secondary caustic from the central one. 

The displacement along the axis that joins the two lenses is the middle point $[\zeta(0)+\zeta(\pi)]/2$ and in our case is

\begin{equation}
    \text{$\zeta_{center}$}=s-\frac{1}{s^n}.
\end{equation}
because the origin of our system is  in the second lens.

Now we want to find out the size of the caustics in the close and wide separation.

For the wide case we have an extension of the caustics in the parallel direction (with respect to the lens axis), given by $[\zeta(0)-\zeta(\pi)]$

\begin{equation}
\Delta\zeta_{||,wide}= 2(n+1)\frac{ \gamma}{s^{\frac{m (n+1)}{m+1}} \left(s^{n+1}-1\right)^{\frac{1}{m+1}}} \label{Deltawide}
\end{equation}

and in the vertical direction, orthogonal to lens axis, $[\zeta(-\pi/2)-\zeta(\pi/2)]$:

\begin{equation}
\Delta\zeta_{\perp,wide}=2(n+1)\frac{ \gamma}{s^{\frac{m (n+1)}{m+1}} \left(s^{n+1}+n\right)^{\frac{1}{m+1}}}
\end{equation}

\begin{figure}[t]
\centering
\includegraphics[height=6.5 cm]{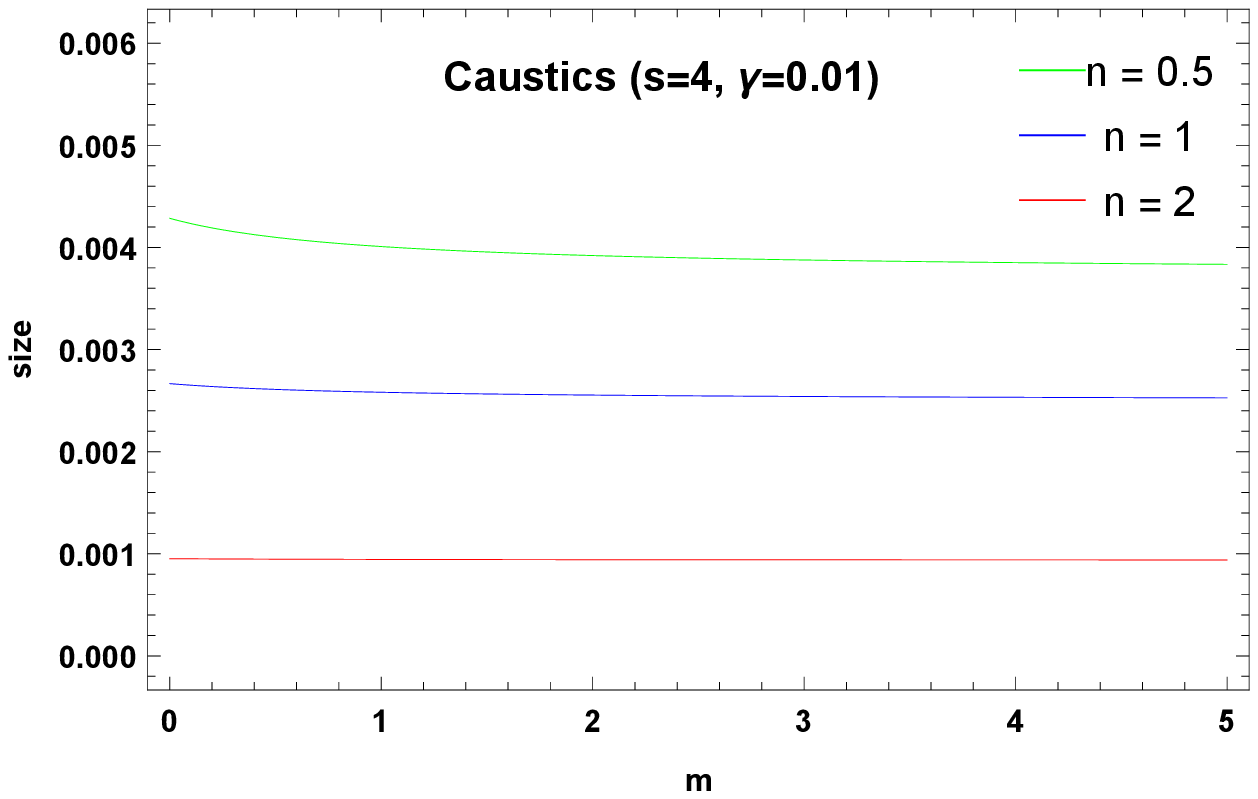}
\includegraphics[height=6.5 cm]{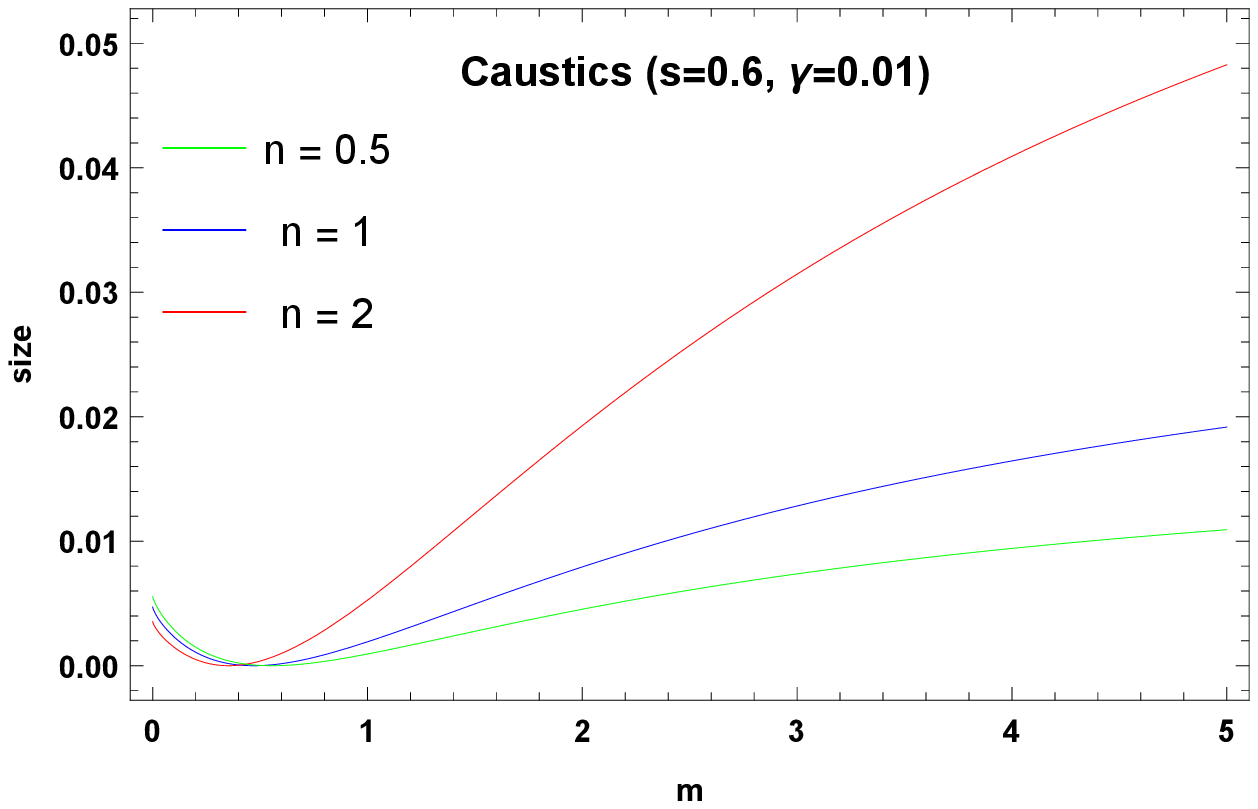}
\caption{Upper panel: size of the caustic in the wide case for $s=4$, $\gamma=0.01$, for three different fixed $n=0.5,1,2$, with variable $m$. Lower panel: size of the caustic in the close case for $s=0.6$, $\gamma=0.01$, for three different fixed $n=0.5,1,2$, with variable $m$.}\label{widesize}
\end{figure}

In Fig.\ref{widesize}, upper panel, we show the size of the caustic for three different fixed $n=0.5,1,2$ with variable $m$. Keeping $\gamma$ and $s$ fixed, the size is almost independent of $m$, as can be seen by neglecting $n$ in the sum in the denominator of Eq. (\ref{Deltawide}).

In the close regime, in order to find the position of the central caustic, we evaluate the lens equation in $\rho_{\pm}$ for $\theta=\pm\pi/2$ and we need to distinguish the case $mn>1$ from the case $mn<1$ changing the sign after taking the square root.

We put $\theta=\pi/2$ and $\rho_+$ in Eq. (\ref{lensGammazro}), we take the imaginary part and we find

\begin{equation}
Im\big[\zeta_+(\pi/2)\big]=\begin{cases}  (m+1)\gamma\bigg[\frac{(1- s^{n+1}) }{m s^{n+1}} \bigg]^{\frac{m}{m+1}}, & \mbox{if }mn<1\mbox{ } \\ (n+1)\gamma \bigg[\frac{1 }{s^{m(n+1)}(n+ s^{n+1})}\bigg]^{\frac{1}{m+1}}, & \mbox{if }mn>1\mbox{}
\end{cases}    
\end{equation}


In order to find the position of the two secondary caustics we need to calculate $\zeta(\pi/2)$ with $\rho_-$, then we take the imaginary part and we find

\begin{equation}
Im\big[\zeta_-(\pi/2)\big]=\begin{cases}
  (n+1)\gamma \bigg[\frac{1 }{s^{m(n+1)}(n+ s^{n+1})}\bigg]^{\frac{1}{m+1}}, & \mbox{if }mn<1\\
(m+1)\gamma\bigg[\frac{(1- s^{n+1}) }{m s^{n+1}} \bigg]^{\frac{m}{m+1}},  & \mbox{if }mn>1
\end{cases}    
\end{equation}



The measure of the transverse size of the secondary caustics is the difference between the last two formulas, $Im\big[\zeta_+(\pi/2)\big]-Im\big[\zeta_-(\pi/2)\big]$ and we plot the result in Fig. \ref{widesize}, lower panel, for $s=0.6$, $\gamma=0.01$, for three different fixed $n=0.5,1,2$, with variable $m$.


We can see that the size increases with $m$ and $n$, coherently with what is found in Ref. \cite{Bozza_2016}, where large values of $m$ and $n$ produce giant triangular caustics in the close regime. We finally stress that, for $mn<1$, and so for $n=0.5$ especially (green line), the two branches exchange role because of the elliptic umbilic catastrophe. Then, we must change the sign in our formula for the size. 

In these approximations the error is given by a term $O(\gamma^{2})$, which means a $1\%$ error for the case $\gamma=0.01$ examined in our plots.

\section{Conclusions}

In this paper we have generalized our previous study of binary lenses with $1/r^n$ potential \cite{Bozza_2016}, by extending it to the case of mixed binaries. Of course the mathematics of this general case is interesting from several points of view, since many earlier results are put in a more general context. However, this case is also important from the astrophysical point of view. In fact, we now have the critical curves and caustics of pairs of galaxies with different halos, or we may apply our results to cases in which one object is made up of exotic matter and the other one is a normal star. For direct applications of our results to astrophysical objects, we remind that all plots are in units of the Einstein radius of the first lens. This can be calculated by standard formulae for any specific lens models.

Our figures, together with those of Ref. \cite{Bozza_2016} may be considered as a complete atlas of critical curves and caustics in binary lensing by $1/r^n$ potentials. We have studied different limits in which the stronger (weaker) lens has a steeper (gentler) potential in all three topology regimes. 

We have shown that an elliptic umbilic catastrophe exists for $mn<1$ and calculated its position. We have also described the pseudocaustic in the $m=0$ limit. We have calculated the boundaries of the three topology regimes and provided analytic approximations for the wide binary and the extremely small-strength secondary lens.

With respect to the $m=n$ binary lens case, we note that for large $m$ we still have large secondary triangular caustics, but they are not as giant as those in Ref. \cite{Bozza_2016}. In fact, the presence of a more standard lens in the system mitigates the behavior at large distances and pushes back these caustics to more normal sizes. Indeed, these structures are quite sensitive to the parameters of the lens.

This fact helps us recall that the mixed binary lens described here is still obtained by the linear superposition of the potentials of two isolated objects. This is physically relevant whenever we can neglect the non-linear terms in Einstein equations. Even when this is not possible, our results may serve as a basis for more accurate calculations.




\reftitle{References}




\externalbibliography{refs.bib}



\end{document}